\documentclass[twocolumn]{aastex631}
\usepackage[fleqn]{amsmath}
\usepackage[caption=false]{subfig}
\usepackage{float}
\usepackage{hyperref}
\usepackage{graphics}

\newcommand{\R}{\textcolor{red}}
\newcommand{\B}{\textcolor{blue}}

\newcommand{\be}{\vspace{.5cm}\begin{equation}}
\newcommand{\angles}{$\left(\theta, \phi, \psi\right)$\,}
\newcommand{\grad}{$^\circ$}
\newcommand{\ellip}{$\varepsilon$\,}

\shorttitle{Deprojection statistics
}
\shortauthors{de Nicola et al.}

\begin{document}

\title{Intrinsic shapes of Brightest Cluster Galaxies}

\author{Stefano de Nicola}\thanks{denicola@mpe.mpg.de}
\affiliation{Max-Planck Institute for Extraterrestrial Physics, Giessenbachstrasse 1, D-85748, Garching (Germany)}

\author{Roberto P. Saglia}
\affiliation{Max-Planck Institute for Extraterrestrial Physics, Giessenbachstrasse 1, D-85748, Garching (Germany)}
\affiliation{Universit{\"a}ts-Sternwarte M{\"u}nchen, Scheinerstrasse 1, D-81679, Munich, Germany}

\author{Jens Thomas}
\affiliation{Max-Planck Institute for Extraterrestrial Physics, Giessenbachstrasse 1, D-85748, Garching (Germany)}
\affiliation{Universit{\"a}ts-Sternwarte M{\"u}nchen, Scheinerstrasse 1, D-81679, Munich, Germany}

\author{Claudia Pulsoni}
\affiliation{Max-Planck Institute for Extraterrestrial Physics, Giessenbachstrasse 1, D-85748, Garching (Germany)}

\author{Matthias Kluge}
\affiliation{Universit{\"a}ts-Sternwarte M{\"u}nchen, Scheinerstrasse 1, D-81679, Munich, Germany}
\affiliation{Max-Planck Institute for Extraterrestrial Physics, Giessenbachstrasse 1, D-85748, Garching (Germany)}

\author{Ralf Bender}
\affiliation{Universit{\"a}ts-Sternwarte M{\"u}nchen, Scheinerstrasse 1, D-81679, Munich, Germany}
\affiliation{Max-Planck Institute for Extraterrestrial Physics, Giessenbachstrasse 1, D-85748, Garching (Germany)}

\author{Lucas M. Valenzuela}
\affiliation{Universit{\"a}ts-Sternwarte M{\"u}nchen, Scheinerstrasse 1, D-81679, Munich, Germany}

\author{Rhea-Silvia Remus}
\affiliation{Universit{\"a}ts-Sternwarte M{\"u}nchen, Scheinerstrasse 1, D-81679, Munich, Germany}



\begin{abstract}

We discuss the statistical distribution of galaxy shapes and viewing
angles under the assumption of triaxiality by deprojecting observed
Surface Brightness (SB) profiles of 56 Brightest Cluster Galaxies
coming from a recently published large deep-photometry sample. For
the first time, we address this issue by directly measuring axis
ratio profiles without limiting ourselves to a statistical analysis
of average ellipticities. We show that these objects are strongly
triaxial, with triaxiality parameters 0.39 $ \leq T \leq $ 0.72, have on average
axis ratios $< p(r) > = $ 0.84 and $< q(r) > =$ 0.68, and are more spherical in the central regions
but flatten out at large radii. Measured shapes in the outskirts agree well with the shapes found for simulated massive galaxies and their dark matter halos from both the IllustrisTNG and the Magneticum simulations, possibly probing the nature of dark matter.
In contrast, both simulations fail to reproduce
the observed inner regions of BCGs, producing too flattened objects.

\end{abstract}



\section{Introduction} \label{Sec_intro}

Massive elliptical galaxies show, when observed projected on the plane
of the sky, smooth elliptical contours with mild boxy deviations
(\,$a_4 \,\rangle\,0$, \citealt{Bender87}), small twists (typically
$\lesssim 10^\circ$) and increasing ellipticity $\left(\varepsilon\right)$
profiles towards outer radii \citep{Goullaud18}. The
average projected flattening is $\langle q' \rangle \sim 0.8$
\citep{Tremblay1996, Weijmans2014, Chen16, Ene18}, with $q' \equiv 1 -
\varepsilon$. The presence of isophote twists, together with other
factors such as the statistical distribution of ellipticity profiles
provides evidence for the triaxiality of these objects
\citep{Illingworth77, Bertola78, Vincent2005}. \\ Various works in the
last 25 years have studied the average statistical
distribution of intrinsic shapes for large galaxy samples at different
redshifts $z$ \citep{Tremblay1996, Vincent2005, Weijmans2014, Chang13, Chen16, Ene18, Li18}.  Deprojecting these distributions
(e.g. \citealt{Tremblay1996}) allows to recover the {\it average}
or {\it typical} intrinsic shape of the galaxies. These studies find
that most massive objects are indeed triaxial, with a mean triaxiality
parameter $T = \left(1 - p^2\right) / \left(1 - q^2\right)$
\citep{Franx91} in the range $\left[0.4, 0.8\right]$
\citep{Vincent2005}, where $p \equiv b/a$, $q \equiv c/a$, and $a \geq
b \geq c$ are the lengths of the three principal axes of the density
ellipsoid. However, no study has yet attempted to \textit{directly
  measure radially resolved intrinsic shapes of individual galaxies in
  large samples}.

In a recent paper, \citet{dN20} have presented a triaxial deprojection
routine that fits the intrinsic shape of ellipsoidal galaxies and
allows to constrain the viewing angles under which an object is seen
by photometric data alone. This can be refined further in combination
with the dynamical modeling of appropriate stellar kinematics (de
Nicola et al., in prep.), since the number of deprojections which need
to be tested is drastically reduced, allowing the study of large
samples of galaxies.  \\ An interesting group of massive galaxies
consists of the so-called Brightest Cluster Galaxies (BCGs). According
to \citet{Matthias20}, a BCG is defined as the closest galaxy to the
geometrical and the kinematical centre of a given galaxy cluster,
although not necessarily the most luminous galaxy of the cluster
itself. Lying deep in the potential well of the cluster, these giant
ellipticals are able to increase their mass through processes such as
galaxy mergers \citep{Contini18}, cannibalism or tidal stripping
\citep{Mo2011}. In a recent paper by \citet{Matthias20}, a sample of 170
BCGs was analyzed in great detail using extremely deep photometric
observations, revealing that BCGs follow different scaling relations
with respect to ordinary ETGs. BCGs are also interesting because
their outer parts have probably grown predominantly by collision-less
accretion and, hence, in a manner similar to the proposed growth of
(collision-less) dark-matter halos.

The first goal of this paper is to constrain the intrinsic shapes and
viewing angles of a representative subsample (56 objects) of this BCG
catalogue with the deprojection method of \citet{dN20}. Then, our
second goal is to compare the recovered shapes to the ones of simulated
massive galaxies and their dark matter halos. For this purpose we
consider the IllustrisTNG \citep{Nelson18, Springel18, Marinacci18,Naiman18, Pillepich18} and Magneticum pathfinder (\citealt{Hirschmann14, Teklu15}\footnote{\url{www.magneticum.org}})
simulations. These cosmological (magneto)-hydrodynamical model the
formation and evolution of galaxies in a $\Lambda$CDM Universe
including recipes for star formation and evolution, chemical
enrichment of the inter-stellar medium, gas cooling and heating, black
hole and supernova feedback. These simulations produce galaxy
populations with properties in reasonable agreement with observations
\citep{Remus17, Teklu17, Genel18, VandeSande19, RodriguezGomez19, Claudia20, Remus21}.


The paper is structured as follows. Section 2 describes the galaxy
sample used in this work. In Section 3 we explain the deprojection
procedure. In Section 4 we present the results on the statistics of
triaxial shapes and compare our findings with the TNG and Magneticum
simulations. Finally, we draw our conclusions in Section 5. Throughout
the paper we assume a flat cosmology with $H_0 = 69.6 km s^{-1}
Mpc^{-1}$ and $\Omega_m = 0.286$.

\section{The sample}  \label{Sec.sample}

The BCGs studied in this work come from a recently published sample
(Tab. 1 of \citealt{Matthias20}). Each BCG was observed in the $g'$
band with the 2m Fraunhofer telescope at the Wendelstein Observatory
(see \citealt{Matthias20} for technical details). The photometry has
exactly the requirements needed for the present work, being extremely
deep (down to $m_{g'} \sim 30$ mag) and reaching very large radii
(typically well beyond 100 kpc). From the complete sample, we extract
those objects for which supplementary F606W HST photometry (typical
resolution $\sim$0.15") is available, excluding galaxies that are
overall unrelaxed (see below). To combine these high-resolution data with those coming from Wendelstein observations, we first select the radii where we have data from both observation sets that are not affected by seeing (typically from 5 to 15 arcsec from the center), then we interpolate HST photometry at Wendelstein radii. Finally, we convert the HST data to the g' band, by determining the sky level and the scaling factor that minimize the differences between the two photometric sets. We combine the two sets by taking the HST values in the inner 10-15 arcsec, and the Wendelstein values at larger radii. In this way we have photometric data with both very high resolution
in the center and also extending out to $\sim$100 kpc for the majority
of the objects\footnote{As shown by \citet{Matthias20}, at larger
radii (typically as SB approaches 27-28 mag arcsec$^{-2}$) isophotal
shape profiles become too noisy to be estimated reliably.}. We
complement this list with further 8 BCGs which we recently observed
both in the H and/or Ks bands at the 8.4m Large Binocular Telescope (LBT)
using Adaptive Optics, with typical resolution of $\sim$0.4". We combine the LBT photometry with the Wendelstein one using the same approach described above. In App.~\ref{App.ref_tests} we show a comparison of the deprojection of two galaxies with and without high-resolution photometry to explore possible photometric effects, showing that they are small. Without high resolution data, the deprojections cannot probe the central regions of the galaxies, but reliable profiles are derived at larger radii. \\
This allows us to add 16 more BCGs with only Wendelstein data, for a total of 56 galaxies. The average isophotal flattening $\langle q'
\rangle$ is $\sim$\,0.77, although almost every BCG becomes very flat
$\left(q' \lesssim 0.4 \right)$ at large radii. In
App.~\ref{App.profiles} we show the $\varepsilon$ and PA profiles for
every BCG of the sample.\\ Since BCGs often show signs of interactions
with other neighbor galaxies of the cluster or AGN activity in the
central regions \citep{Matthias20}, and given that our triaxial code
works under the assumption of (smooth) "deformed ellipsoids" (see
eq. 29 of \citealt{dN20}), we omit the
innermost/outermost isophotes from the deprojection when we find signs of incomplete relaxation (for example, in the form of bumpy \ellip or PA profiles). This happens in
the very center (e.g. AGNs, ongoing accretion) or in the very outer
parts (where dynamical time scales are large).  Notes on individual
galaxies can be found in App.~\ref{App.notes}. Since we are interested
in comparing our findings with simulations at large radii, we
try, when possible, to extend the deprojection up to $2-4 R_e$, with the
values for half-light radii $R_e$ taken from Tab. 4 of \citet{Matthias20}.

\begin{figure*}[ht!]
\centering
\includegraphics[scale=.6]{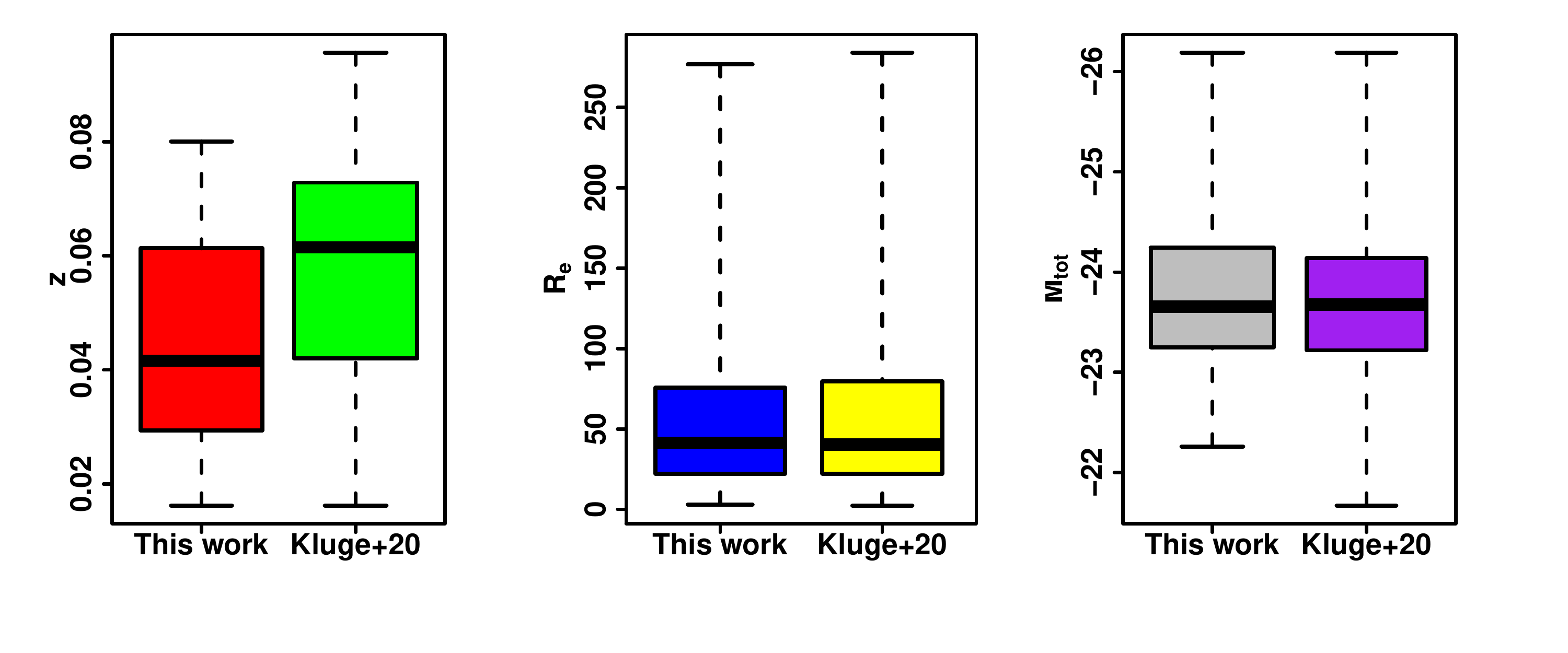}

\caption{Boxplots showing redshifts (left panel), effective radii (central panel) and $g'$ magnitudes (right panel) for \citet{Matthias20}'s sample and our sub-sample. Our sub-sample is
biased towards galaxies at lower redshift and also slightly towards galaxies with high R$_\text{e}$. Boxes are drawn from the first to the third quartiles, with the horizontal line in the middle of the boxes denoting the median. Whiskers span across the whole data range.}
\label{Fig.boxplot}
\end{figure*}

\subsection{Selection effects}

The full sample of \citet{Matthias20} is drawn from the Abell–Corwin–Olowin
catalog (ACO, \citealt{Abell89}) by adopting redshift and
volume-limiting constraints. It contains BCGs with redshift $z
\lesssim 0.08$ (with 15 outliers) and has a slight Malmquist bias (see
their Figure 3). However, a comparison of this sample with other large
samples, also drawn from ACO, shows that they have about 80-90\% of the objects
in common. \\ In order to check the completeness of our sub-sample, we define
R${_\text{N}}$ as the ratio of the number of galaxies in our sub-sample
N$_{\text{sub}}$ to the number of galaxies in the full sample
N$_{\text{full}}$. We become progressively incomplete at larger
redshifts: at $z \leq 0.04$, we have $0.5 \leq \text{R}_\text{N} \leq
1.0$, but at larger redshifts R${_\text{N}}$ is $\leq 0.2$.
Moreover, Fig.~\ref{Fig.boxplot}, left panel, shows that the mean
redshift and redshift range covered by our subsample are smaller than
the ones of the parent sample. This is expected, since both HST- and
LBT-observed galaxies are at lower redshift than the average, and we picked them
up to perform dynamical modeling.\\
We do not find significant selection effects when
considering the size (Fig.~\ref{Fig.boxplot}, middle panel) or the total absolute magnitude of the galaxies
M$_{\text{tot}}$, (Fig.~\ref{Fig.boxplot}, right panel). The results are
summarised in Tab.~\ref{Tab.selection}.

\begin{table}
\begin{tabular}{ccc}
Variable & Intervals & R${_\text{N}}$ \\
\hline \hline
 & $\leq 0.02$ & 1.0 \\
$z$ & $\left[0.02,0.05\right]$ & $\sim$0.5 \\
 & $\geq 0.05$ & $\sim$0.15 \\
 & & \\
R$_e$ & $\geq 50$ & $\sim$0.25 \\
  & & \\
M$_{\text{tot}}$ & $\left[-22,-26\right]$ & $\sim$0.25 \\
\hline
\end{tabular}
\vspace{.5cm}

\caption{ The ratio R${_\text{N}}$ of the number of galaxies in our
  sub-sample N$_{\text{sub}}$ for a given interval to the number of
  galaxies in the full sample N$_{\text{full}}$ for three variables of
  interest. Our sub-sample is biased towards low-redshift galaxies, while no significant bias in magnitude and effective radii is
  found.}
\label{Tab.selection}
\end{table}

\section{Deprojection procedure}

In this section we describe the deprojection parameters used for the
BCGs. An extensive description of the deprojection routine itself is
given by \citet{dN20}. In short, given the observed surface luminosity
$I_{\text{obs}}=L/pc^2$ onto a polar elliptical grid, the code
searches for the three-dimensional luminosity density $\rho$, placed
onto an ellipsoidal grid, whose corresponding projected surface
luminosity $I_{\text{fit}}$ minimizes RMS$=\sqrt{\langle \left(\ln (
  I_{\text{obs}}/I_{\text{fit}})\right)^2 \rangle}$. The algorithm
works under the assumption that a galaxy can be described by what we
call a “deformed ellipsoid”, namely an ellipsoid whose radius is given
by

    \begin{equation}
        m^{2-\xi(x)} = x^{2-\xi(x)} + \left[\frac{y}{p(x)}\right]^{2-\xi(x)} + \left[\frac{z}{q(x)}\right]^{2-\xi(x)}
        \label{eq.def_ellips}
    \end{equation}

\noindent where the exponent $\xi$ can be used to generate disky \newline ($\xi \rangle 0$) or
boxy ($\xi \langle 0$) bias. The three one-dimensional functions $p(x), q(x)$,
and $\xi(x)$, along with the density on the x-axis $\rho_x (x)$, specify $\rho$ at each point of the grid. Finally, the code uses a one-dimensional radial smoothing on $p(x), q(x)$,
$\xi(x)$ and $\rho_x (x)$ to penalize against unsmooth solutions.

\subsection{Choice of parameters} \label{Sec.pars_depro}

Table~\ref{Tab.parameters} highlights for each BCG the
parameters used for the deprojections. In the following we will give a detailed 
explanation of these parameters.

\begin{itemize}
    \item \textit{Grid sampling.} It is important to place both the
      observed $I_{\text{obs}}$ and deprojected $\rho$ on grids large
      enough to reproduce photometric information properly, but at the
      same time a very large grid would slow down the code
      significantly. Therefore, we start by interpolating the SB onto
      a finely sampled grid, and then gradually reduce both the number
      of radial $\left(n_{r'}\right)$ and angular points
      $\left(n_{\theta'}\right)$ as long as the comparison with the
      observations remains acceptable, that is, an error below 1\% for
      every photometric variable (SB, $\varepsilon$, PA, a$_4$,
      a$_6$). Then, we set the number of radial points of the
      $\rho$-grid $n_{r} = n_{r'} + 20$, while the numbers of angular
      points $n_{\theta}$ and $n_{\phi}$ are typically the same as
      $\left(n_{\theta'}\right)$ or slightly larger.
    
    \item \textit{Grid extension in radius.} The innermost radii are
      the same for both grids, namely the semiminor axis of the
      innermost isophote. These have to be estimated for every BCG by
      taking into account the spatial resolution of the observations
      and by checking whether the galaxy shows central activity. In
      this last case, central regions are omitted. The outermost radii
      for the SB grid are estimated by making use of our software for
      isophotal fitting \citep{Bender87}. We typically stop when we
      the isophotal shape profiles become noisy and have to be set to a
      constant values (typically at SB$\sim$27-28 mag asec$^{-2}$). The
      largest radii of the $\rho$-grid are then a few times those of
      the corresponding SB grid.
    
    \item \textit{Grid flattenings.} The flattening of the SB grid can
      easily be estimated by considering the isophote PAs. For
      example, if a galaxy shows isophote structures with the major
      axes aligned along the vertical axis on the plane of the sky
      $y'$, it makes sense to use an elliptical polar grid flattened
      in the $x'$-direction, even with a 15-20$^\circ$ twist. As far
      as it concerns $\rho$, we first assume a spherical grid, then we
      re-deproject the galaxies at the best-fit inclination(s) using
      the recovered $\langle p(r) \rangle$ and $\langle q(r) \rangle$
      as flattenings.
    
    \item \textit{Smoothing.} As shown in de Nicola et al. (in prep.)
      we can recover the true \textit{intrinsic} density of a
      triaxial $N$-body simulation with an RMS of $\sim$10\%. Since it is not
      entirely clear how to estimate the smoothing a priori, we take
      the four $\lambda$-values $\left[\lambda_\rho, \lambda_p, \lambda_q, \lambda_\xi\right]$ (cfr. eq. 30 of \citealt{dN20}) used
      with the simulation divided by a factor of 2, to take into
      account that our data are less noisy than the $N$-body
      simulation (the smoothing scales as $\lambda^{-2}$). Since the smoothing value affects the RMS one gets
      at the end (the higher the smoothing, the higher the RMS), we
      verified that for the best galaxies the RMS was comparable to
      the one we got for the simulation. The values we chose are [0.6,0.03,0.03,0.3]. A more rigorous
      implementation would be the minimization of the Akaike
      Information Criterion (AIC, \citealt{Akaike74}), as shown by
      \citet{Mathias21, Jens22}. 
      We defer this to a forthcoming paper.
    
    \item \textit{Constraints on $p, q$.} Our code allows the
      possibility of deprojecting by imposing constraints on $p,
      q$. Since the code has shown excellent results in terms of
      recovering the right profiles, we only impose $p, q \geq 0.2$,
      to prevent too flat solutions which may give problems to the
      fit.
    
\end{itemize}


\begin{table*}
\begin{center}
\resizebox{1.5\columnwidth}{.4\paperheight}{%
\begin{tabular}{ccccccc}
Galaxy & arcsec/kpc & r$_{\text{min}}$ & r$_{\text{max}}$ & $I_{\text{obs}}$ grid & $\rho$\,grid & Photometry \\
\hline \hline
2MASXJ0753 & 1.17 & 2.13 & 128.1 & 40 $\times$ 15 & 60 $\times$ 16 $\times$ 16 & L + W\\
2MASXJ0900 & 1.426 & 2.35 & 104.2 & 50 $\times$ 15 & 70 $\times$ 15 $\times$ 15 & W\\
2MASXJ1358 & 1.225 & 1.98 & 74.04 & 40 $\times$ 10 & 60 $\times$ 11 $\times$ 11 & W\\
IC613 & 0.653 & 0.0940 & 116.1 & 50 $\times$ 10 & 70 $\times$ 11 $\times$ 11 & H + W\\
IC664 & 0.679 & 0.0984 & 76.0 & 40 $\times$ 10 & 60 $\times$ 11 $\times$ 11 & H + W\\
IC1101 & 1.48 & 1.06 & 70.2 & 40 $\times$ 12 & 60 $\times$ 13 $\times$ 13 & H + W\\
IC1565 & 0.765 & 0.101 & 243.2 & 40 $\times$ 10 & 60 $\times$ 11 $\times$ 11 & H + W\\
IC1634 & 1.336 & 1.58 & 156.8 & 50 $\times$ 10 & 70 $\times$ 11 $\times$ 11 & W\\
IC1695 & 0.987 & 0.150 & 149.1 & 40 $\times$ 10 & 60 $\times$ 11 $\times$ 11 & H + W\\
IC1733 & 0.714 & 0.528 & 33.8 & 30 $\times$ 12 & 50 $\times$ 13 $\times$ 13 & W\\
IC2378 & 0.990 & 0.762 & 58.5 & 40 $\times$ 10 & 60 $\times$ 11 $\times$ 11 & H + W\\
IC5338 & 1.10 & 4.91 & 178.3 & 40 $\times$ 10 & 60 $\times$ 11 $\times$ 11 & H + W\\
LEDA1518 & 1.248 & 4.16 & 117.1 & 40 $\times$ 10 & 60 $\times$ 11 $\times$ 11 & W\\
LEDA2098 & 1.467 & 2.82 & 127.3 & 40 $\times$ 10 & 60 $\times$ 11 $\times$ 11 & W\\
MCG+01-60 & 1.16 & 0.622 & 38.4 & 30 $\times$ 12 & 50 $\times$ 13 $\times$ 13 & W\\
MCG-02-02 & 1.083 & 3.44 & 173.7 & 40 $\times$ 10 & 60 $\times$ 11 $\times$ 11 & W\\
MCG+02-04 & 0.869 & 0.213 & 184.3 & 50 $\times$ 12 & 70 $\times$ 13 $\times$ 13 & H + W \\
MCG+02-27 & 0.653 & 0.114 & 103.4 & 40 $\times$ 10 & 60 $\times$ 11 $\times$ 11 & H + W\\
MCG+02-58 & 1.52 & 1.69 & 100.1 & 40 $\times$ 10 & 60 $\times$ 11 $\times$ 11 & H + W \\
MCG+03-04 & 1.375 & 2.76 & 66.34 & 40 $\times$ 10 & 60 $\times$ 11 $\times$ 11 & W\\
MCG+03-38 & 0.886 & 0.165 & 87.6 & 40 $\times$ 10 & 60 $\times$ 11 $\times$ 11 & H + W\\
MCG+04-28 & 2.53 & 1.15 & 299.7 & 40 $\times$ 12 & 60 $\times$ 13 $\times$ 13 & H + W \\
MCG+05-32 & 1.44 & 1.19 & 312.9 & 40 $\times$ 10 & 60 $\times$ 11 $\times$ 11 & H + W \\
MCG+05-33 & 1.23 & 2.64 & 67.1 &40 $\times$ 10 & 60 $\times$ 11 $\times$ 11 & H + W \\
MCG+09-13 & 1.362 & 3.78 & 103.1 & 40 $\times$ 10 & 60 $\times$ 11 $\times$ 11 & W\\
MCG+09-20 & 1.296 & 2.49 & 133.4 & 40 $\times$ 10 & 60 $\times$ 11 $\times$ 11 & W\\
NGC708 & 0.332 & 0.0551 & 61.2 & 40 $\times$ 15 & 60 $\times$ 15 $\times$ 15 & H + W\\
NGC910 & 0.354 & 0.611 & 70.4 & 40 $\times$ 10 & 60 $\times$ 11 $\times$ 11 & H + W\\
NGC1128 & 0.486 & 0.0895 & 31.7 & 40 $\times$ 10 & 60 $\times$ 11 $\times$ 11 & H + W \\
NGC1129 & 0.361 & 0.160 & 98.0 & 60 $\times$ 12 & 80 $\times$ 12 $\times$ 12 & H + W\\
NGC1275 & 0.359 & 0.739 & 81.9 & 40 $\times$ 10 & 60 $\times$ 11 $\times$ 11 & W\\
NGC2329 & 0.396 & 0.534 & 43.3 & 50 $\times$ 12 & 70 $\times$ 13 $\times$ 13 & H + W\\
NGC2804 & 0.559 & 1.58 & 35.6 & 40 $\times$ 10 & 60 $\times$ 11 $\times$ 11 & W\\
NGC3550 & 0.703 & 0.505 & 99.9 & 40 $\times$ 12 & 60 $\times$ 13 $\times$ 13 & L + W\\
NCG3551 & 0.640 & 0.382 & 33.4 & 30 $\times$ 12 & 50 $\times$ 13 $\times$ 13 & W\\
NGC4104 & 0.577 & 0.970 & 60.2 & 40 $\times$ 12 & 60 $\times$ 13 $\times$ 13 & L + W\\
NGC4874 & 0.469 & 0.0905 & 98.6 & 50 $\times$ 10 & 70 $\times$ 11 $\times$ 11 & H + W\\
NGC6166 & 0.622 & 0.718 & 94.9 & 40 $\times$ 10 & 60 $\times$ 11 $\times$ 11 & H + W\\
NGC6173 & 0.592 & 0.0613 & 109.2 & 40 $\times$ 10 & 60 $\times$ 11 $\times$ 11 & H + W\\
NGC6338 & 0.552 & 0.334 & 132.4 & 50 $\times$ 10 & 70 $\times$ 11 $\times$ 11 & H + W\\
NGC7647 & 0.818 & 0.323 & 97.5 & 50 $\times$ 10 & 70 $\times$ 11 $\times$ 11 & H + W\\
NGC7649 & 0.835 & 0.659 & 94.1 & 40 $\times$ 10 & 60 $\times$ 11 $\times$ 11 & H + W\\
NGC7720 & 0.611 & 0.720 & 153.1 & 40 $\times$ 10 & 60 $\times$ 11 $\times$ 11 & H + W\\
NGC7768 & 0.545 & 0.307 & 102.3 & 40 $\times$ 10 & 60 $\times$ 11 $\times$ 11 & H + W\\
SDSSJ0837 & 2.67 & 1.66 & 118.0 & 40 $\times$ 20 & 60 $\times$ 20 $\times$ 20 & L + W\\
UGC716 & 1.19 & 0.832 & 175.7 & 40 $\times$ 10 & 60 $\times$ 11 $\times$ 11 & L + W\\
UGC727 & 1.135 & 1.89 & 114.4 & 40 $\times$ 10 & 60 $\times$ 11 $\times$ 11 & W\\
UGC1191 & 1.21 & 0.638 & 124.5 & 40 $\times$ 20 & 60 $\times$ 20 $\times$ 20 & L + W\\
UGC2232 & 0.958 & 0.152 & 95.3 & 40 $\times$ 10 & 60 $\times$ 11 $\times$ 11 & H + W\\
UGC2413 & 0.690 & 0.379 & 113.4 & 50 $\times$ 10 & 70 $\times$ 11 $\times$ 11 & H + W\\
UGC4289 & 0.587 & 0.587 & 50.1 & 40 $\times$ 10 & 60 $\times$ 11 $\times$ 11 & H + W\\
UGC6394 & 0.847 & 1.99 & 97.2 & 40 $\times$ 10 & 60 $\times$ 11 $\times$ 11 & W\\
UGC9799 & 0.691 & 4.13 & 97.7 & 40 $\times$ 10 & 60 $\times$ 11 $\times$ 11 & H + W\\
UGC10143 & 0.708 & 0.584 & 110.6 & 50 $\times$ 10 & 70 $\times$ 11 $\times$ 11 & H + W\\
UGC10726 & 1.15 & 0.704 & 189.8 & 60 $\times$ 12 & 80 $\times$ 12 $\times$ 12 & L + W\\
VV16IC & 0.354 & 0.401 & 181.8 & 40 $\times$ 10 & 60 $\times$ 11 $\times$ 11 & L + W\\
\hline
\end{tabular}
}
\end{center}
\vspace{.5cm}
\caption{
\textit{Col. 1}: Galaxy name. \textit{Col. 2}: arcsec/kpc conversion factor. \textit{Cols. 3-4}: Smallest and largest isophotal radii, in kpc. \textit{Cols. 5-6}: $I_{\text{obs}}$ and $\rho$-grid dimensions. \textit{Col. 7}: The available photometry (W: Wendelstein, H: HST, L: LBT).}
\label{Tab.parameters}
\end{table*}

\subsection{Viewing angles} \label{Ssec.va}

Deprojections yield in general a non-unique solution \citep{Gerhard96,
  VDB97, dN20}, unless one uses constraints about the galaxy shape. We
impose the resulting deprojection to consist of a series of concentric
perfect ellipsoids, i.e. setting $\xi(r) = 0$ at all radii\footnote{This assumption is justified by the fact that for the BCGs the deviations from elliptical shapes are small, i.e. $|a_4/a| < 5$\% in the deprojected regions.}, and
without imposing biasing towards certain $p, q$ profiles.  Further
parameters to be considered are the three viewing angles
$\left(\theta, \phi, \psi\right)$ required to identify the orientation
in space of a triaxial galaxy. The first two give the orientation of
the line-of-sight (LOS) in space, whereas the third one is a rotation
about the LOS itself. If one could measure these angles, then the
intrinsic axis ratios of the ellipsoid could be calculated
analytically. Since this is not usually the case, we need to sample a
grid of viewing angles and deproject every BCG trying out every set of
viewing angles. The assumption of triaxiality, i.e. the galaxy has an
8-fold symmetry, allows us to sample the two viewing angles
$\left(\theta, \phi\right)$ in $\left[0, \pi/2\right]$, while $\psi$
needs to be sampled in $\left[0, \pi\right]$. We sample each angle in
10\grad step, which gives us a total of 1800 inclinations to test for
each BCG. As shown by \citet{dN20}, although the viewing angle
estimate through deprojections is not always perfect, the list of
"good" solutions (see below) \textit{always} includes the correct
viewing angles. \\ In the same paper, the authors show that sampling
one octant does not guarantee the "canonical" $1 \geq p \geq q$ order
relationship (cfr. their Table 2). However, for a given deprojected
density at a given set of viewing angles for which this does not
happen, \textit{it is always possible to find another set of viewing
  angles for which the deprojection is equivalent but with $1 \geq p
  \geq q$}, possibly in the octant with $\phi \in \left[\pi/2,
  \pi\right]$. In this case, if the density happens to be a "good" one
in terms of the RMS (see below), we re-perform the deprojection for
this new set of viewing angles such that the inequality $1 \geq p \geq
q$ holds. \\ In order to find the "good" deprojections, we isolate for
every galaxy $g$ all viewing angles for which the deprojection has
$\mathcal{R}_g \leq \delta \times \mathcal{R}_{\text{min,g}}$, where
$\mathcal{R}_g \equiv \text{RMS}_g$ is the RMS and
$\mathcal{R}_{\text{min,g}}$ is the smallest RMS that we find among
all viewing angles that we sampled for galaxy $g$. The factor $\delta$
determines how stringent the cut-off is. In another work aiming at
modeling an $N$-body simulation (de Nicola et al., in prep.), we adopt
values in the range $\delta = 1.2-1.5$ studying the impact of the choice of this value on the number of plausible deprojections. Here, we try to be conservative adopting
$\delta = 1.5$ The resulting light densities are those that we use to
derive the results shown in Sec.~\ref{Sec.shape_results}. \\
Finally, we note that there may still be a degeneracy between the model and orientation parameters, but this is small and will not be taken into account throughout the remainder of the paper. In App.~\ref{App.ref_tests} we examine the case of NGC7647, and derive profiles for the best-fit viewing angles, stopping the deprojection when the RMS reaches 1.5 RMS$_{min}$ (which is the best value achievable). 
These p \& q profiles are well within the range obtained by considering all possible viewing angles.

\begin{figure*}[h!]

\subfloat{\includegraphics[width=.3\linewidth]{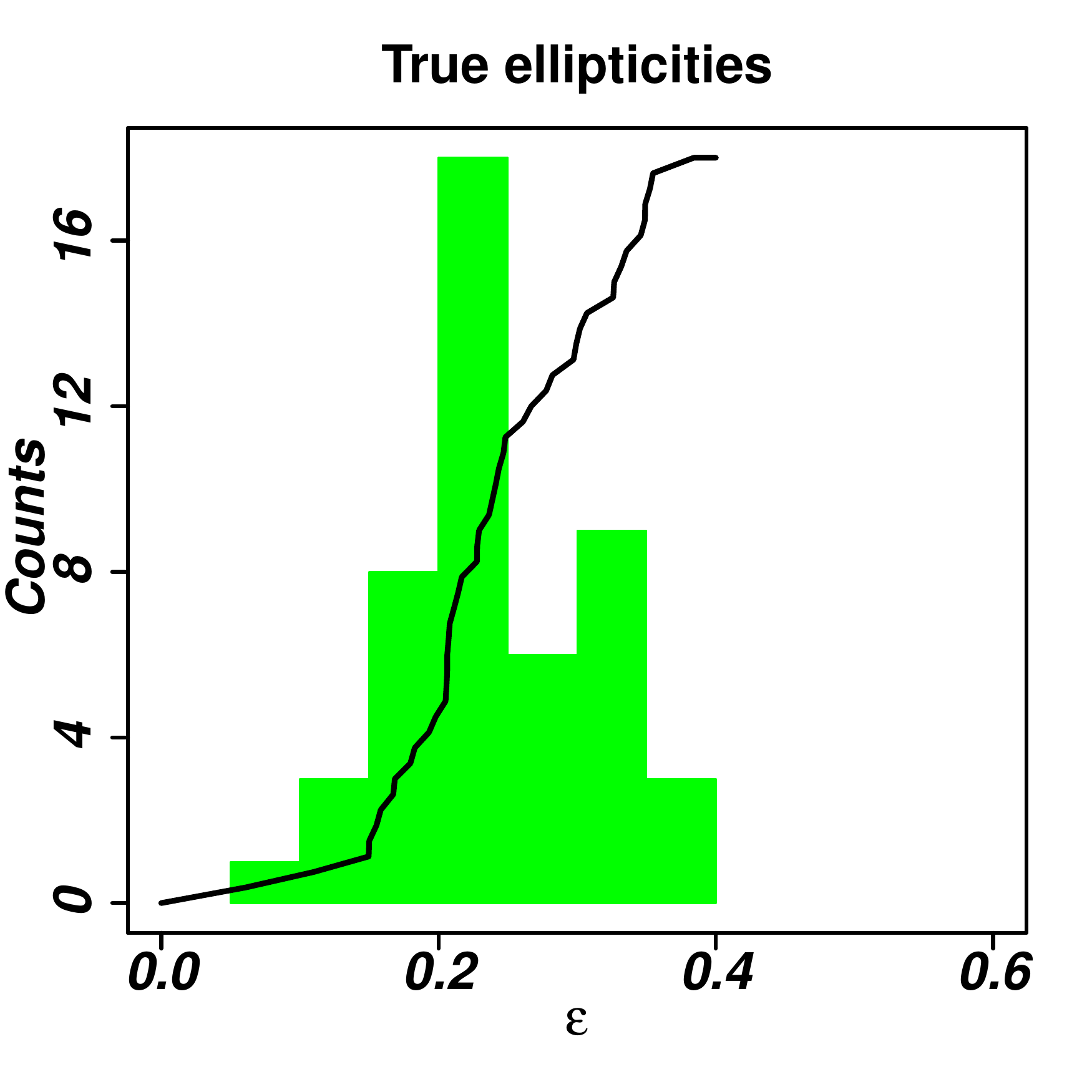}}
\subfloat{\includegraphics[width=.3\linewidth]{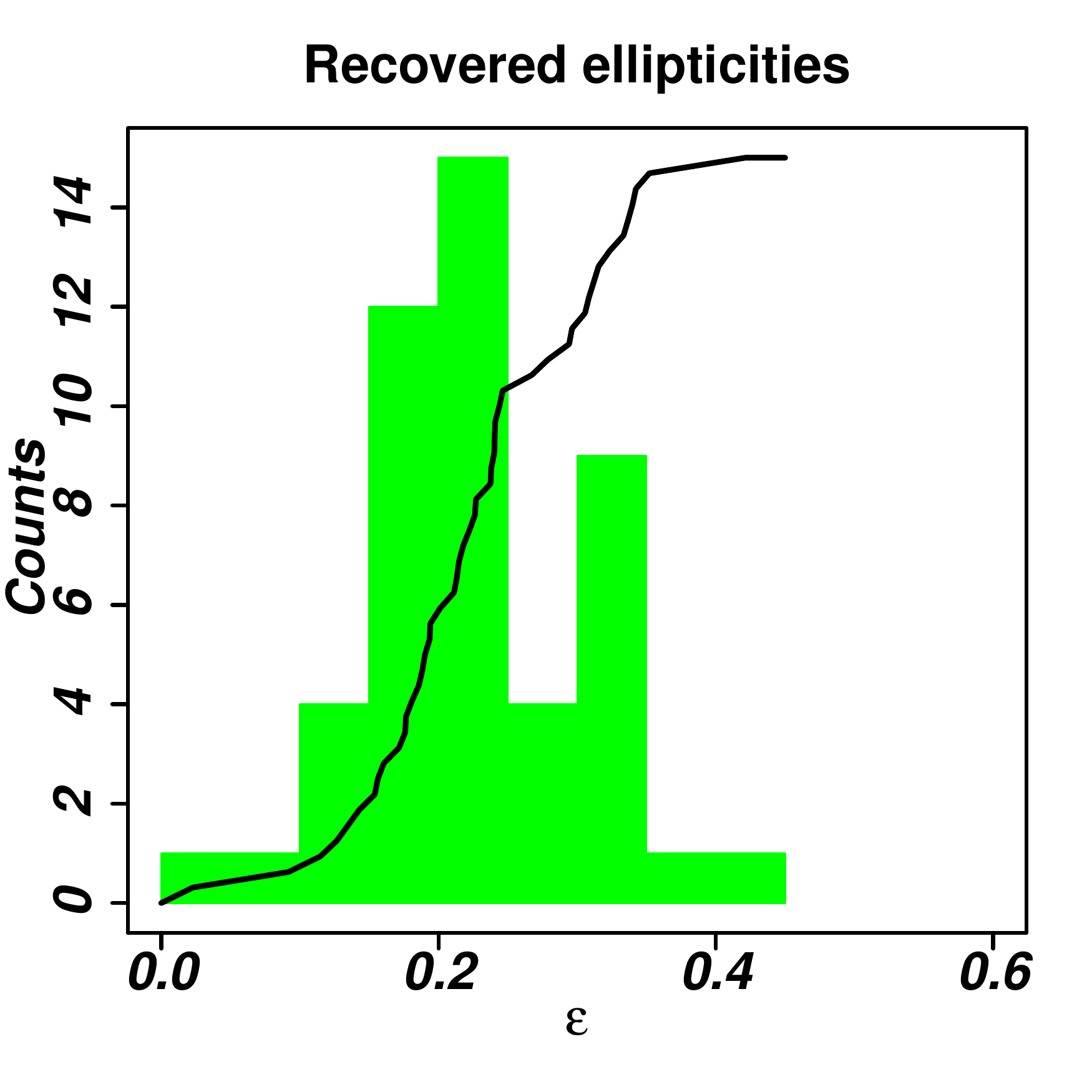}}
\subfloat{\includegraphics[width=.3\linewidth]{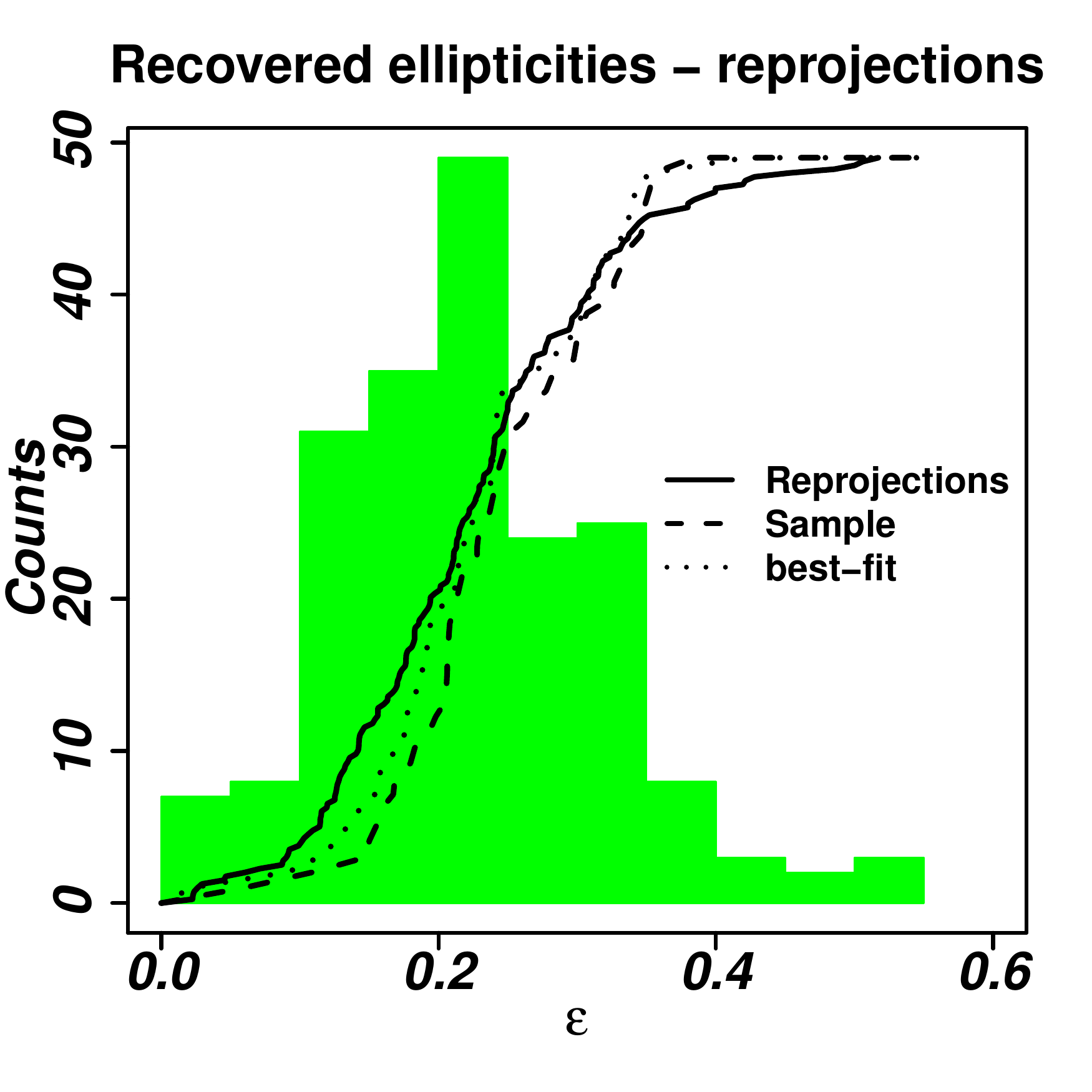}}

\subfloat{\includegraphics[width=.3\linewidth]{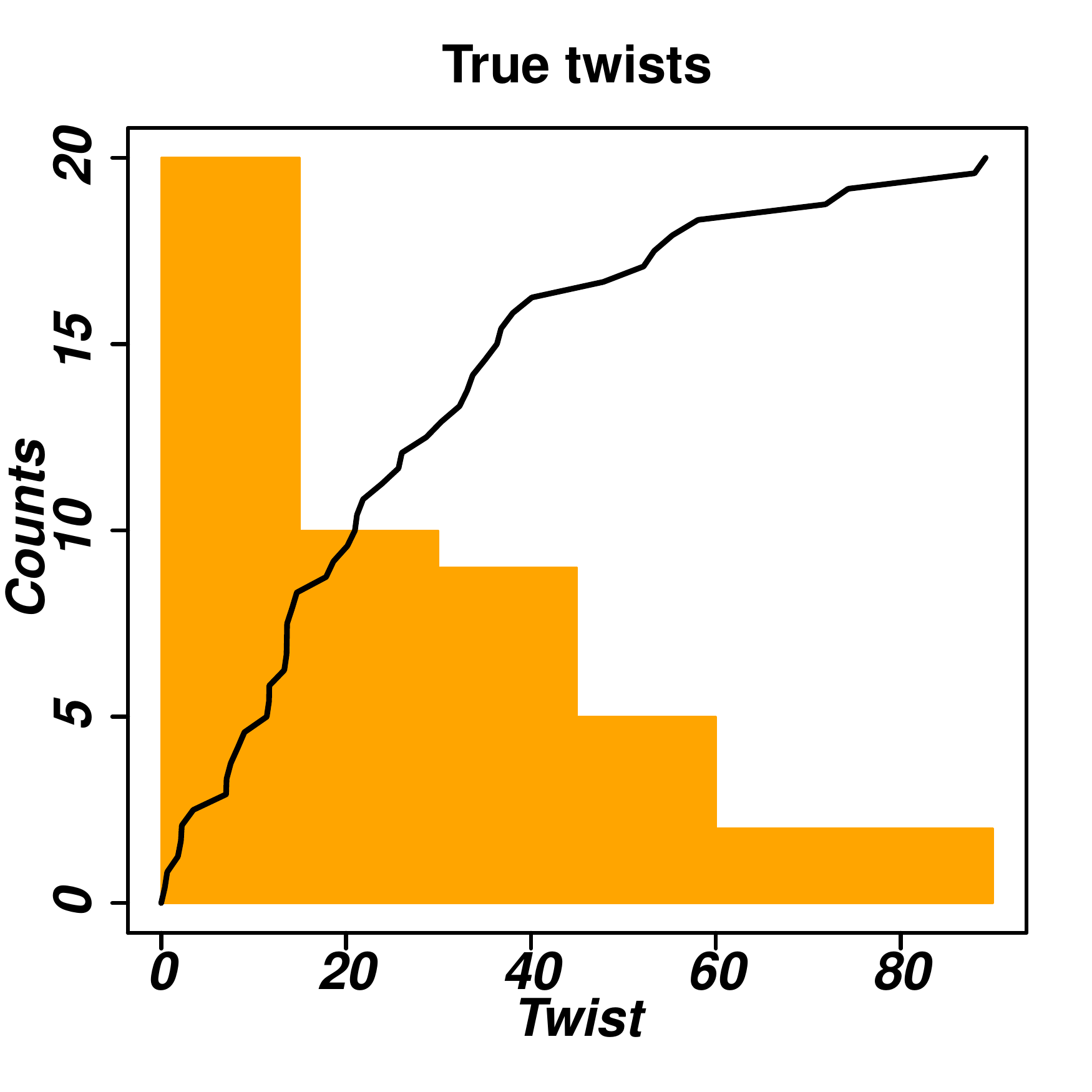}}
\subfloat{\includegraphics[width=.3\linewidth]{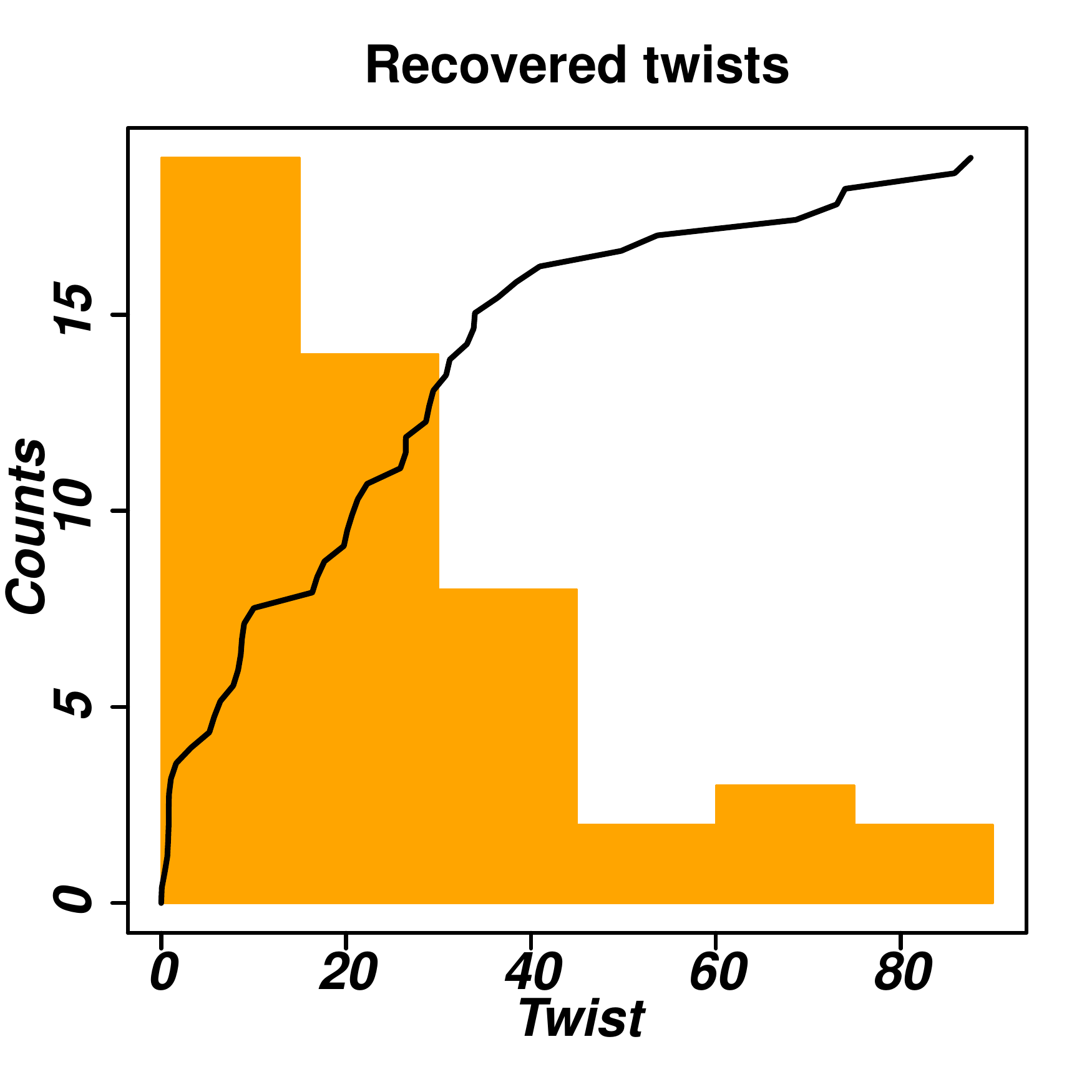}}
\subfloat{\includegraphics[width=.3\linewidth]{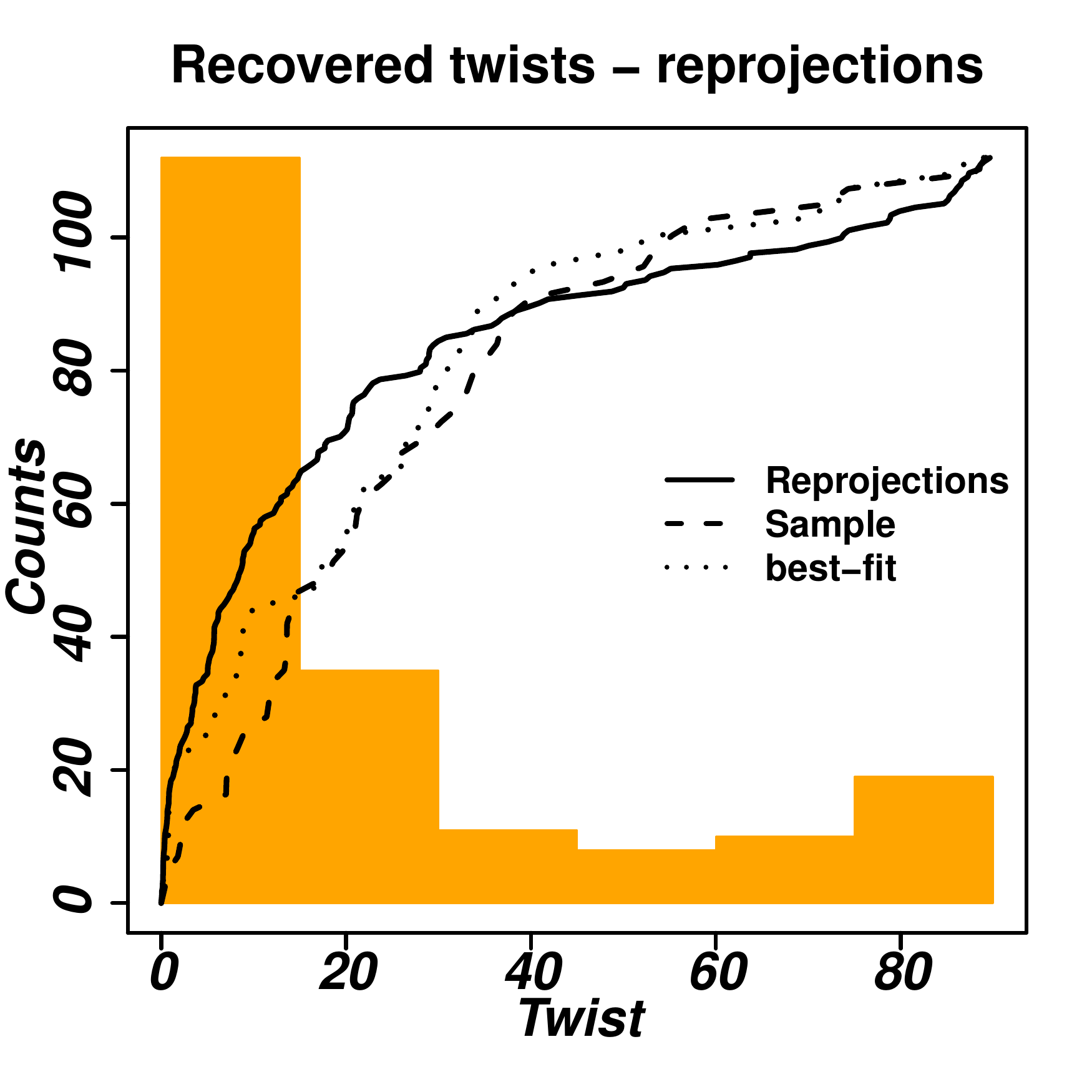}}

    \caption{For both \ellip (top row) and PA (bottom row) we present
      three plots. The left panels show the observed mean values, the
      central panels the recovered values for the best-fit angles and
      the right panels the values obtained by re-projecting the
      best-fit solutions at random viewing angles. The solid lines are
      the cumulative distribution functions (cdfs), rescaled to the
      maximum count values. On the right panels we also show the cdf
      from the left and middle panels by dashed and dotted lines, respectively.}
    \label{Fig.check_photometry}
\end{figure*}

\begin{figure*}
\subfloat{\includegraphics[width=.5\linewidth]{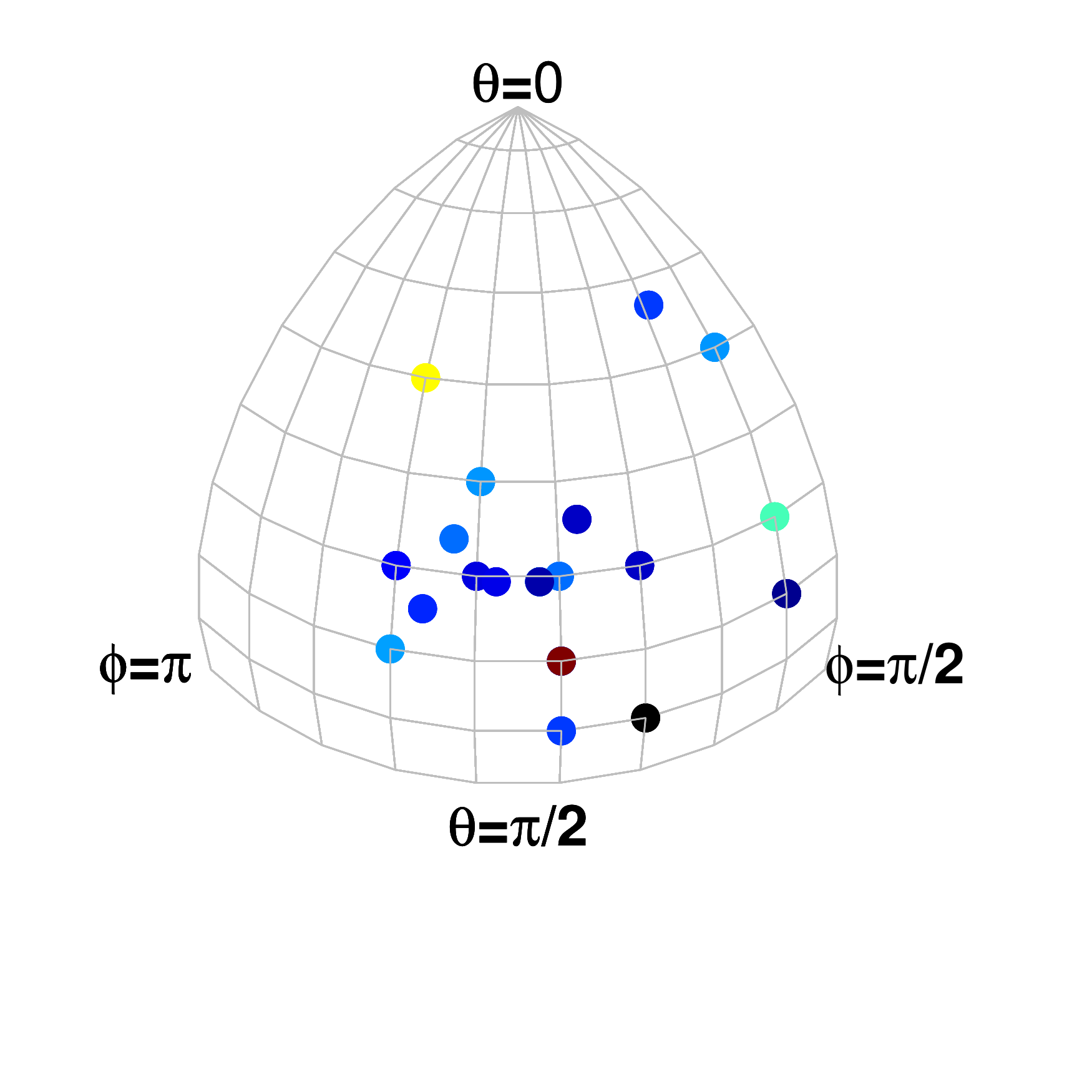}}
\subfloat{\includegraphics[width=.5\linewidth]{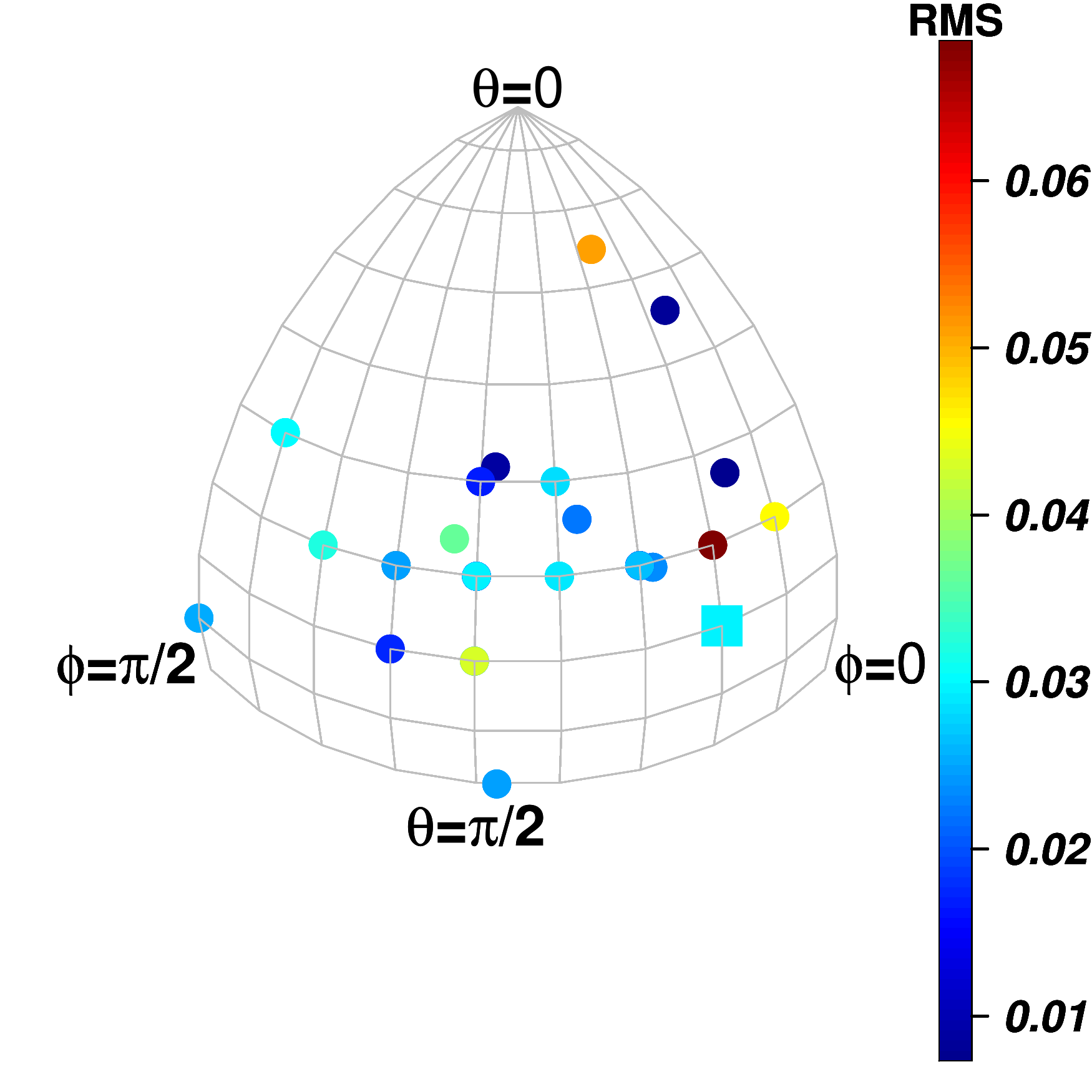}}

\centering
\subfloat{\includegraphics[width=.5\linewidth]{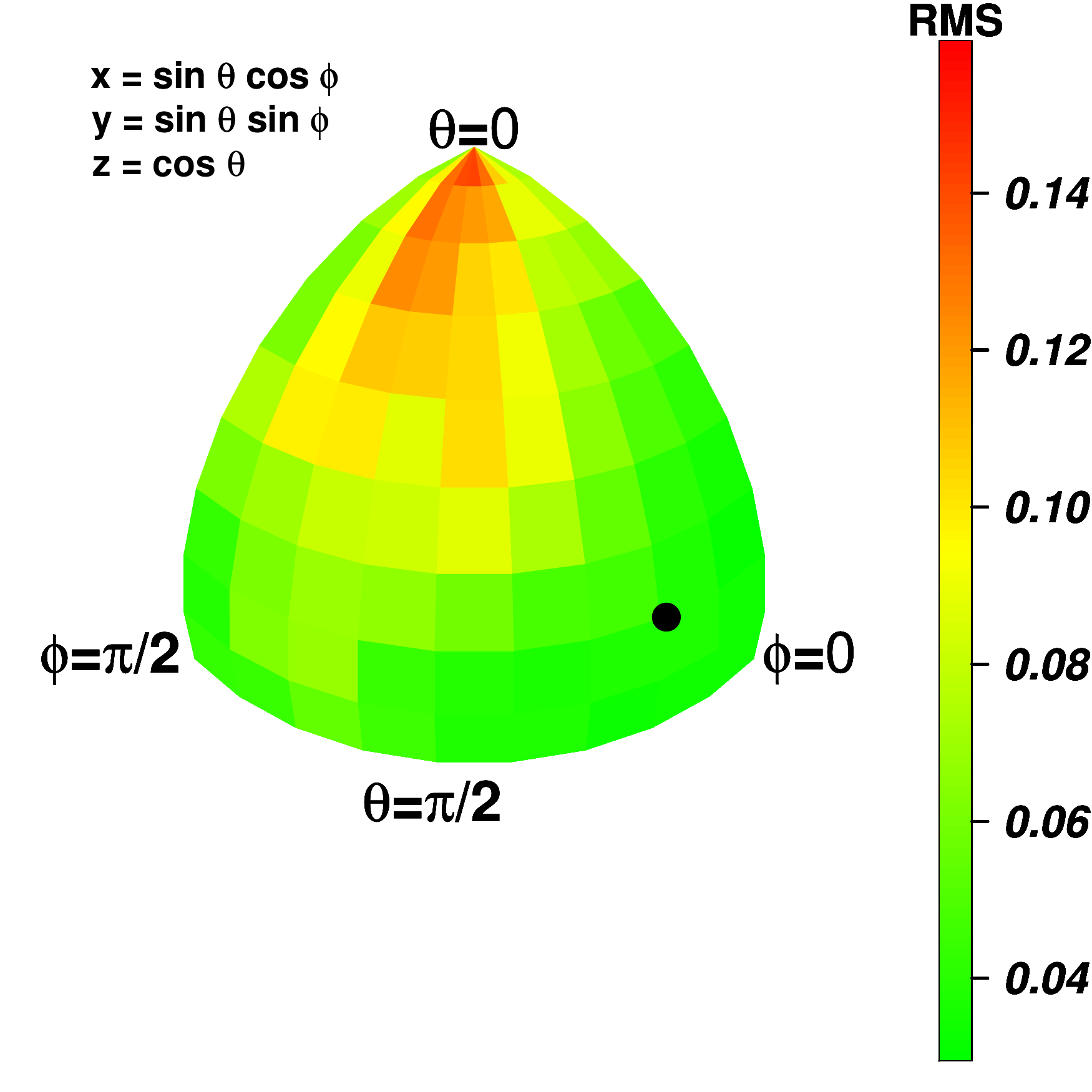}}

\caption{\textit{Top:} Distribution of the best-fit angles on the two
  octants. The square at $\left(\theta, \phi\right) = (70,20)^\circ$
  labels the galaxy NGC708 (see below). The following galaxies appear
  superimposed: UGC 9799, UGC 2413, MCG-02-02 and NGC 2329 with
  $\left(\theta, \phi\right) = (60,50)^\circ$; the galaxies MCG+09-13, IC1634, IC 5338 and IC 664 with $\left(\theta, \phi\right) = (60,40)^\circ$; 
the galaxies 2MASXJ1358 and IC1695 with $\left(\theta, \phi\right) = (50,40)^\circ$; the galaxies LEDA2098, NGC4874 and UGC727 with $\left(\theta, \phi\right) = (60,30)^\circ$; the galaxies LEDA1518 and NGC6338 with $\left(\theta, \phi\right) = (60,60)^\circ$; the galaxies UGC4289 and MCG+03-04 with $\left(\theta, \phi\right) = (60,150)^\circ$; the galaxies MCG+02-04 and NGC 4104 with $\left(\theta, \phi\right) =
  (70,130)^\circ$. \textit{Bottom:} RMS distribution as a function of
  $\left(\theta, \phi\right)$ for the galaxy NGC708, clearly showing
  the goodness of solutions close to the principal axes. The black
  point labels the best-fit solution, located at $\left(\theta,
  \phi\right) = (70,20)^\circ$.}
    \label{Fig.angles_octant}
\end{figure*}


\section{Results}   

Using the cut-off described in Sec.~\ref{Ssec.va} we are able to
reduce the number of inclinations from the initial value of 1800 by at
least a factor of 3. The typical RMS values for the best-fit solutions
are 0.01 to 0.03. The results are summarized in
Tab.~\ref{Tab.depro_results}.

\begin{table*}
\begin{center}
\resizebox{1.5\columnwidth}{.4\paperheight}{%
\begin{tabular}{cccccccc}
Galaxy & $10^2\times$ RMS$_{\text{best}}$ & Good Deprojections & best-fit angles & $\langle p \rangle$ & $\langle q \rangle$ & $\Delta_p$ & $\Delta_q$\\
\hline \hline
2MASXJ0753 & 5.1 & 96 & (26,27,37)$^\circ$ & 0.804 & 0.661 & 0.048 & 0.054\\
2MASXJ0900 & 2.2 & 114 & (54,37,122)$^\circ$ & 0.902 & 0.769 & 0.053 & 0.047\\
2MASXJ1358 & 5.0 & 62 & (50,40,60)$^\circ$ & 0.806 & 0.568 & 0.123 & 0.111\\
IC613 & 2.6 & 250 & (70,50,90)$^\circ$ & 0.879 & 0.782 & 0.047 & 0.040\\
IC664 & 3.2 & 204 & (60,40,145)$^\circ$ & 0.871 & 0.704 & 0.080 & 0.073\\
IC1101 & 3.0 & 28 & (50,80,160)$^\circ$ & 0.798 & 0.644 & 0.076 & 0.075\\
IC1565 & 2.4 & 312 & (60,28,167)$^\circ$ & 0.833 & 0.697 & 0.041 & 0.041\\
IC1634 & 2.8 & 378 & (60,40,110)$^\circ$ & 0.897 & 0.760 & 0.050 & 0.058\\
IC1695 & 2.9 & 376 & (50,40,80)$^\circ$ & 0.854 & 0.712 & 0.061 & 0.057\\
IC1733 & 0.73 & 380 & (53,16,165)$^\circ$ & 0.815 & 0.695 & 0.108 & 0.116\\
IC2378 & 1.6 & 118 & (70,100,150)$^\circ$ & 0.841 & 0.708 & 0.082 & 0.065\\
IC5338 & 3.2 & 58 & (60,40,160)$^\circ$ & 0.824 & 0.615 & 0.082 & 0.073\\
LEDA1518 & 4.1 & 146 & (60,60,120)$^\circ$ & 0.876 & 0.711 & 0.091 & 0.113\\
LEDA2098 & 2.3 & 64 & (60,30,60)$^\circ$ & 0.858 & 0.716 & 0.067 & 0.066\\
MCG+01-60 & 0.81 & 330 & (36,34,110)$^\circ$ & 0.909 & 0.824 & 0.084 & 0.122\\
MCG-02-02 & 2.2 & 94 & (60,50,15)$^\circ$ & 0.786 & 0.560 & 0.084 & 0.081\\
MCG+02-04 & 2.6 & 128 & (70,130,90)$^\circ$ & 0.876 & 0.747 & 0.033 & 0.033\\
MCG+02-27 & 3.0 & 184 & (60,132,3)$^\circ$ & 0.799 & 0.567 & 0.076 & 0.062\\
MCG+02-58 & 4.2 & 96 & (60,100,130)$^\circ$ & 0.763 & 0.632 & 0.082 & 0.101\\
MCG+03-04 & 2.3 & 114 & (60,150,90)$^\circ$ & 0.963 & 0.900 & 0.049 & 0.044\\
MCG+03-38 & 2.1 & 222 & (60,140,145)$^\circ$ & 0.903 & 0.813 & 0.075 & 0.060\\
MCG+04-28 & 12.7 & 144 & (80,120,10)$^\circ$ & 0.877 & 0.763 & 0.107 & 0.127\\
MCG+05-32 & 3.0 & 142 & (56,143,3)$^\circ$ & 0.841 & 0.639 & 0.056 & 0.062\\
MCG+05-33 & 2.6 & 112 & (80,90,165)$^\circ$ & 0.889 & 0.793 & 0.067 & 0.069\\
MCG+09-13 & 2.8 & 66 & (60,40,75)$^\circ$ & 0.822 & 0.611 & 0.083 & 0.088\\
MCG+09-20 & 5.3 & 76 & (40,150,100)$^\circ$ & 0.789 & 0.649 & 0.088 & 0.100\\
NGC708 & 2.7 & 162 & (70,20,130)$^\circ$ & 0.885 & 0.695 & 0.030 & 0.031\\
NGC910 & 3.2 & 142 & (60,70,145)$^\circ$ & 0.877 & 0.747 & 0.078 & 0.103\\
NGC1128 & 2.7 & 216 & (80,130,145)$^\circ$ & 0.941 & 0.884 & 0.015 & 0.021\\
NGC1129 & 4.6 & 494 & (60,10,0)$^\circ$ & 0.888 & 0.780 & 0.047 & 0.044\\
NGC1275 & 3.3 & 80 & (50,140,80)$^\circ$ & 0.780 & 0.599 & 0.087 & 0.083\\
NGC2329 & 1.8 & 140 & (60,50,0)$^\circ$ & 0.930 & 0.848 & 0.034 & 0.031\\
NGC2804 & 1.8 & 32 & (61,132,93)$^\circ$ & 0.897 & 0.772 & 0.068 & 0.046\\
NGC3550 & 6.8 & 422 & (60,20,20)$^\circ$ & 0.973 & 0.937 & 0.028 & 0.050\\
NCG3551 & 0.88 & 114 & (48,48,151)$^\circ$ & 0.856 & 0.687 & 0.090 & 0.093\\
NGC4104 & 1.5 & 74 & (70,130,135)$^\circ$ & 0.587 & 0.290 & 0.083 & 0.082\\
NGC4874 & 1.5 & 302 & (60,30,150)$^\circ$ & 0.925 & 0.818 & 0.025 & 0.025\\
NGC6166 & 2.5 & 80 & (90,47,150)$^\circ$ & 0.824 & 0.588 & 0.082 & 0.063\\
NGC6173 & 1.4 & 56 & (60,120,30)$^\circ$ & 0.724 & 0.427 & 0.088 & 0.096\\
NGC6338 & 2.1 & 66 & (60,60,165)$^\circ$ & 0.816 & 0.630 & 0.055 & 0.045\\
NGC7647 & 3.7 & 60 & (56,53,18)$^\circ$ & 0.773 & 0.623 & 0.033 & 0.032\\
NGC7649 & 3.0 & 96 & (60,130,95)$^\circ$ & 0.784 & 0.507 & 0.081 & 0.079\\
NGC7720 & 1.8 & 130 & (70,60,150)$^\circ$ & 0.753 & 0.502 & 0.076 & 0.050\\
NGC7768 & 2.9 & 114 & (60,40,150)$^\circ$ & 0.732 & 0.515 & 0.098 & 0.080\\
SDSSJ0837 & 2.6 & 364 & (64,136,36)$^\circ$ & 0.798 & 0.543 & 0.101 & 0.075\\
UGC716 & 3.2 & 130 & (40,100,0)$^\circ$ & 0.817 & 0.582 & 0.078 & 0.081\\
UGC727 & 2.6 & 102 & (60,30,170)$^\circ$ & 0.817 & 0.638 & 0.067 & 0.079\\
UGC1191 & 2.5 & 238 & (60,60,40)$^\circ$ & 0.850 & 0.691 & 0.061 & 0.049\\
UGC2232 & 2.2 & 198 & (60,138,62)$^\circ$ & 0.827 & 0.717 & 0.032 & 0.031\\
UGC2413 & 2.2 & 72 & (60,50,130)$^\circ$ & 0.788 & 0.538 & 0.078 & 0.058\\
UGC4289 & 2.3 & 94 & (60,150,60)$^\circ$ & 0.873 & 0.770 & 0.055 & 0.050\\
UGC6394 & 2.9 & 50 & (54,127,102)$^\circ$ & 0.855 & 0.689 & 0.091 & 0.078\\
UGC9799 & 3.0 & 66 & (60,50,145)$^\circ$ & 0.816 & 0.667 & 0.067 & 0.060\\
UGC10143 & 3.3 & 136 & (70,150,165)$^\circ$ & 0.782 & 0.569 & 0.073 & 0.061\\
UGC10726 & 1.7 & 134 & (50,50,80)$^\circ$ & 0.925 & 0.867 & 0.026 & 0.022\\
VV16IC & 3.2 & 142 & (70,50,0)$^\circ$ & 0.738 & 0.514 & 0.066 & 0.046\\
\hline
\end{tabular}
}
\end{center}
\vspace{.5cm}
\caption{
\textit{Col. 1}: Galaxy name. \textit{Col. 2}: Smallest RMS. \textit{Col. 3}: Number of "good" deprojections. \textit{Col. 4}: The best-fit viewing angles. \textit{Cols 5-6}: Average $p$ and $q$ values. \textit{Cols 7-8}: RMS on average $p$ and $q$ values.}
\label{Tab.depro_results}
\end{table*}

\subsection{Reliability of the deprojections}

As a first step we verify that our deprojections do reproduce the
average photometry of the sample. First, we calculate for every galaxy
the mean $\varepsilon$ and the twist, defined as $\Delta \text{PA} =
\text{max(PA) - min(PA)}$, both for the observed and the recovered
photometry. Moreover, we reproject the best-fit densities $\rho_g$ for
every galaxy $g$ at three different random viewing angles, computing
the same averages as above. This is a good test to statistically verify that the recovered intrinsic shapes are compatible with the observed shape distribution. In Fig.~\ref{Fig.check_photometry} we show
the histograms for $\varepsilon$ (top row) and the twist (bottom
row). A Kolmogoroff-Smirnov (KS, \citealt{Kolmogorov33, Smirnov39})
test returns p-values above the canonical 5\% threshold\footnote{This
  corresponds to values in the range 0.194-0.243 for the KS statistics
  D$_n$.} for both the \ellip and PA distributions, with this being
valid for both the best-fit angles and the reprojections at random
viewing angles. This confirms that the recovered photometric variables
are statistically representative of the BCG sample. \\ A second step,
we check the distribution of the two best-fit angles $\langle(\theta,
\phi)\rangle$, which specify the LOS position on the plane of the
sky. In the upper panels of Fig.~\ref{Fig.angles_octant} we plot the two
octants with the best-fit $\langle(\theta, \phi)\rangle$ onto them. We
see that there is a lack of solutions near the principal axes, but
this is only given by the fact that we are plotting \textit{only the
  best-fit solution for each galaxy}. This clearly disfavours such
viewing angles, because isophotal twists cannot occur along the
principal axes of an ellipsoidal body. Hence, fits along an assumed
LOS that coincides with one of the principal axes will deliver larger
values of the RMS. In the bottom panel of Fig.~\ref{Fig.angles_octant}
we show the entire distribution of deprojections over the two octants
for the galaxy NGC708. Solutions on the principal axes are not
excluded but lead on average to less good fits. Other examples are
provided in the notes in App.~\ref{App.notes}.

\subsection{Distribution of intrinsic axis ratios}    \label{Sec.shape_results}

\begin{figure*}[ht!]
    \resizebox{.3\linewidth}{!}{\includegraphics[scale=.4]{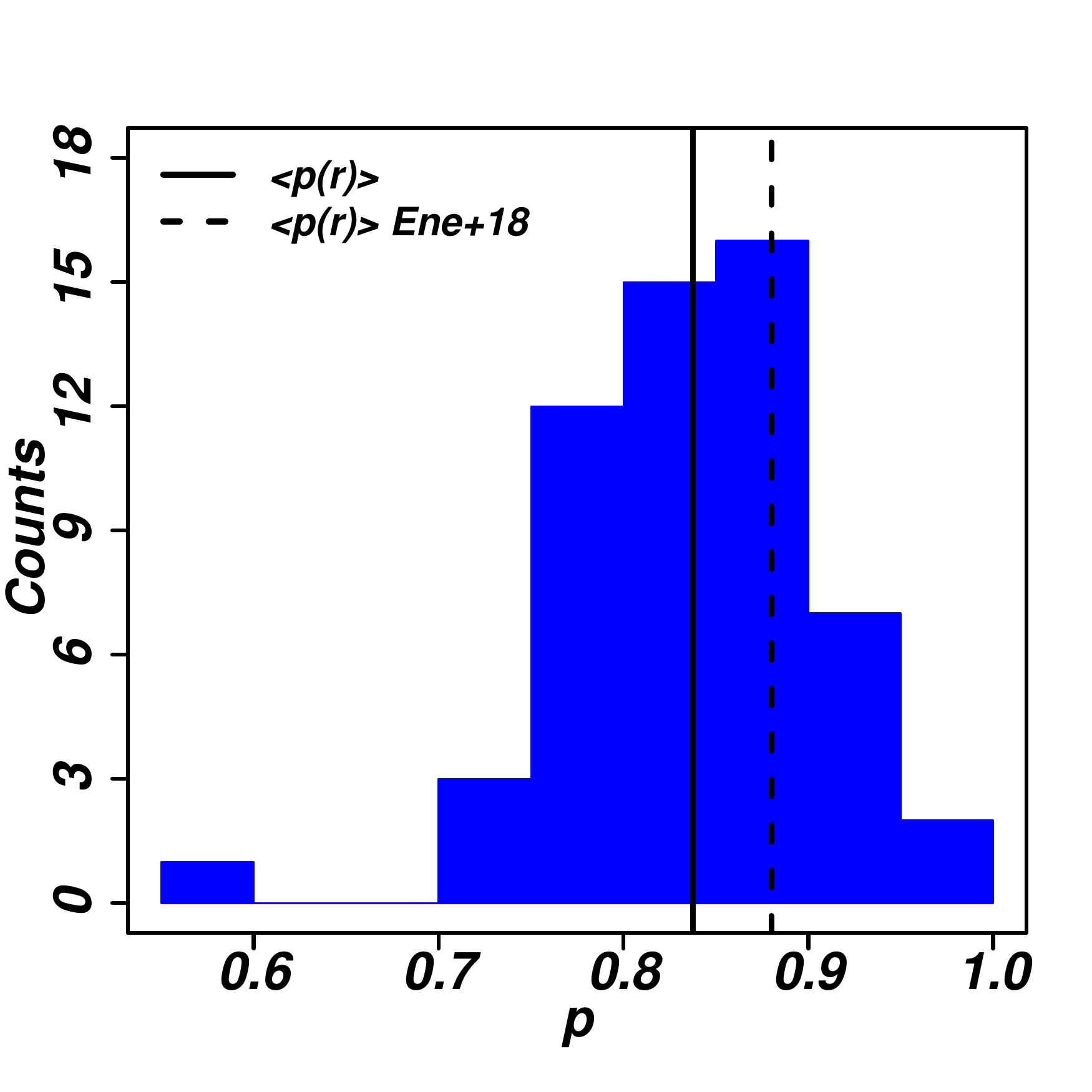}}
\resizebox{.3\linewidth}{!}{\includegraphics[scale=.4]{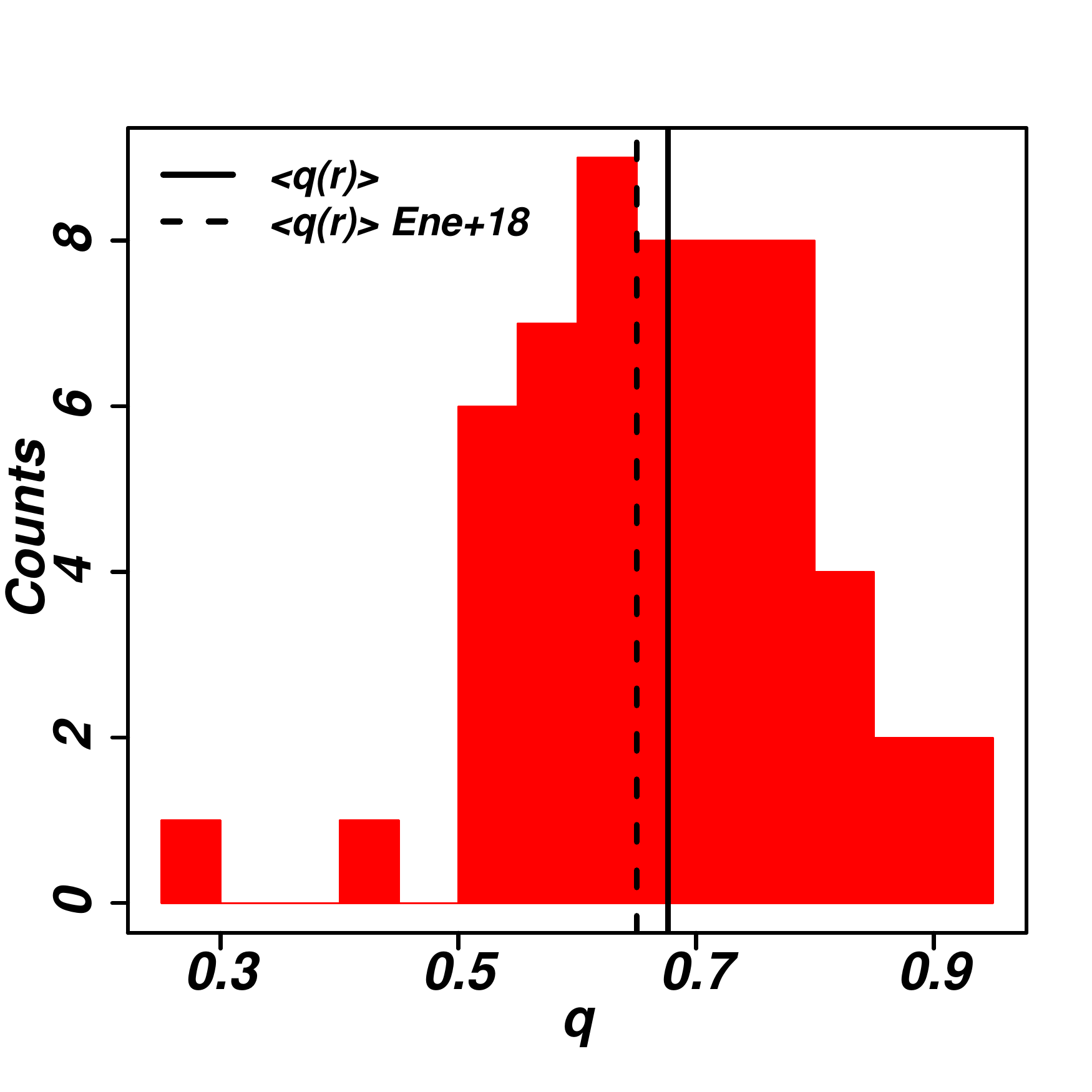}}
\resizebox{.3\linewidth}{!}{\includegraphics[scale=.4]{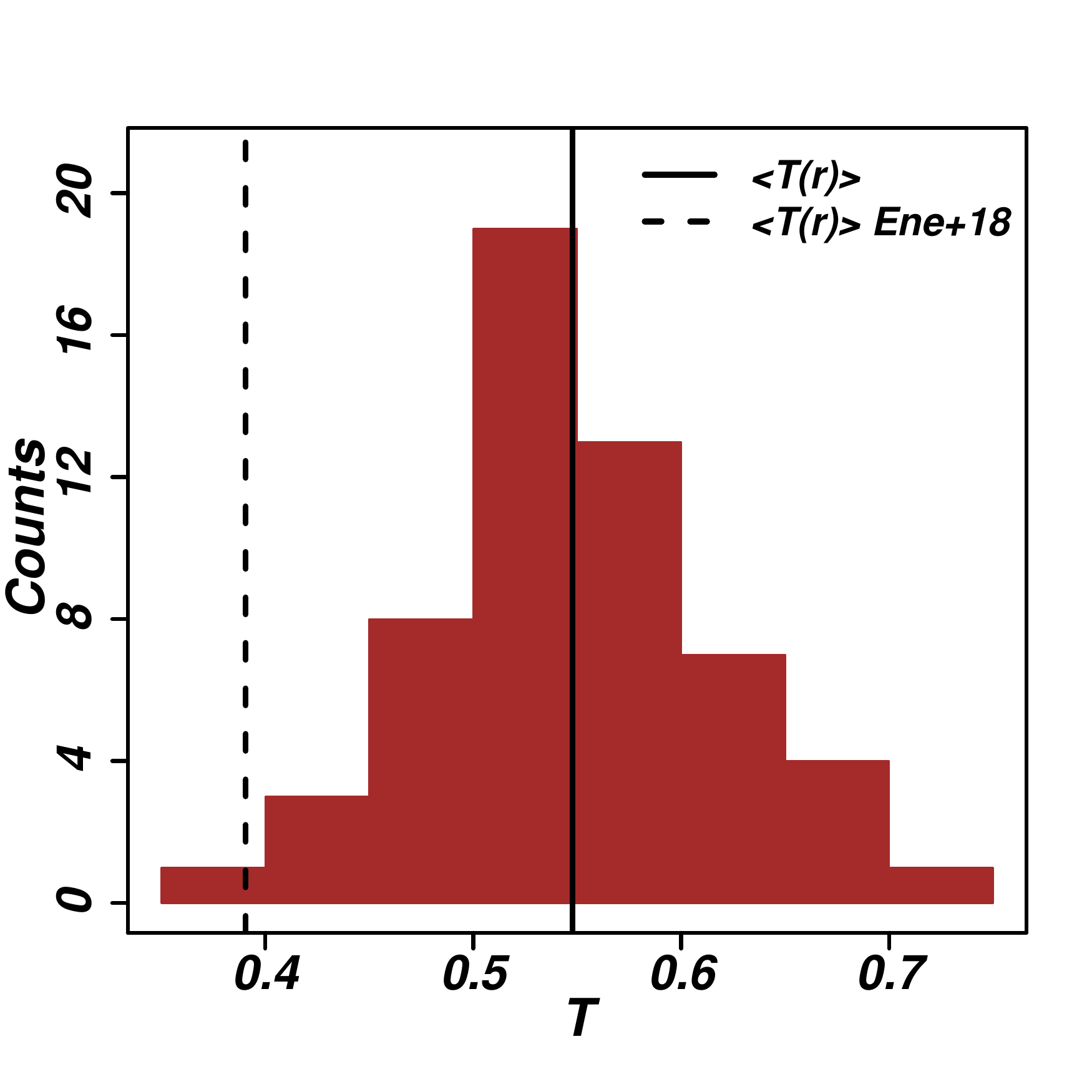}}
\caption{Histograms of the mean intrinsic axis ratios $\langle p(r) \rangle$ (left panel) and $\langle q(r) \rangle$ (middle panel) and of the corresponding triaxiality parameter $\langle T(r) \rangle$ (right panel) for every galaxy of our sample. We also compare the mean values from our histograms with the findings of \citet{Ene18} from the MASSIVE survey for slow rotators.}
\label{Fig.pqTmean}
\end{figure*}

We now present the \textit{measured} shapes of the BCGs. Our
deprojection algorithm \textit{directly} yields the intrinsic axis
ratios $p(r)$ and $q(r)$ as a function of the distance from the galaxy
center. From these profiles, we compute the triaxiality parameter as a
function of the radius $T(r)$. The profiles which we obtain by averaging over all good deprojections are shown in Fig.~\ref{Fig.pqT_BCGs}.\\ 
The left and central panels of
Fig.~\ref{Fig.pqTmean} show the histograms of the average over all radial bins and over all acceptable deprojections $p(r)$ and
$q(r)$ for every galaxy of the sample, i.e. averaged on every "good"
deprojection. We get $\langle p(r) \rangle = $ 0.84 and $\langle q(r)
\rangle = $ 0.68, with scatters of $\sim$0.1 (see
Tab.~\ref{Tab.depro_results} for the values we get for each BCG). For
comparison, \citet{Ene18} used ellipticity distribution and found
$\langle p(r) \rangle = 0.88$ and $\langle q(r) \rangle = 0.65$ for a
sample of slow rotators. The histogram of the mean triaxiality
parameter, presented in the right panel of Fig.~\ref{Fig.pqTmean},
shows that although BCGs follow different scaling relations from
ordinary ETGs, they have 0.39 $ \leq \langle T \rangle \leq$ 0.72, in
agreement with the findings of \citet{Vincent2005} for a sample
consisting only of ordinary ETGs. The conclusion here is
that the triaxiality is extremely high for every object of the sample,
with no object showing a mean triaxiality outside of the
$\left[0.39-0.72\right]$ interval.\\ 
We do not detect correlations
between $\langle p(r) \rangle $, $\langle q(r) \rangle$ or $\langle
T(r) \rangle$ and the size. A weak correlation is seen with absolute magnitudes: bright BCGs appear rounder than fainter ones, having approximately the same triaxiality. The trend is more clearly seen when considering the radial profiles (see Figs.~\ref{Fig.TNG_stars} \& ~\ref{Fig.TNG_DM}). 

\subsection{Comparison with the TNG and Magneticum simulations}

In order to compare the recovered shape profiles to the shape profiles
of simulated galaxies, we use the IllustrisTNG and Magneticum
simulations. We consider the $110.7^3$ Mpc$^3$ and $68^3$ Mpc$^3$
cosmological volumes, respectively, as a good compromise between
resolution and number of massive galaxies simulated. In TNG100, the
mean mass of the stellar particles is $1.4 \times
10^6\,\text{M}_\odot{}$ while the dark matter particles have masses
$7.5 \times 10^6\,\text{M}_\odot{}$. The Plummer equivalent
gravitational softening length for both stars and dark matter at
redshift $z=0$ is $\text{r}_\text{soft} = 0.74$ kpc. Instead,
Magneticum has stellar particles with masses of $2.6 \times
10^6\,\text{M}_\odot{}$ and DM particles with masses of $5.1 \times
10^7\,\text{M}_\odot{}$, while $\text{r}_\text{soft} = 2$ kpc for DM
and 1 kpc for stars, respectively. \\ We select simulated galaxies
with total mass larger than $10^{13} M_\odot$ that are the most
massive members of their group (so-called 'central'). We divide these
galaxies into 2 mass bins, with the number of objects in each mass bin
summarized in Tab~\ref{Tab.TNG}. From these, we derive $p(r)$ and
$q(r)$ profiles for the dark matter and the stellar component
separately. This is done by diagonalizing the inertia tensor

\begin{equation}
    I_{ij} = \frac{\sum_n m_{n} x_{n,i} x_{n,j}}{\sum_n m_{n}},
\label{eq:inertia_tensor}
\end{equation}

\noindent where $x_{n,i}$ are the coordinates of the stellar particles and $m_n$ their mass, calculated in ellipsoidal shells \citep{Zemp11}. We choose 10 radial bins logarithmically spaced along the intrinsic major axis of the
galaxies from 3 to 100 kpc.  In each step, the iterative procedure
adjusts the flattening of the ellipsoidal shell and the direction of
principal axes to the iso-density contours, until it converges within
1\% in both $p$ and $q$. We verified that the variation in the
direction of the principal axes are generally within 5 degrees between
3 and 100 kpc and that fixing their position to a mean direction (for
example, measured within 1 effective radius) slightly overestimates
the axis ratios by a few percent, up to a median $\sim3\%$ in $p$ and
$7\%$ in $q$ at 100 kpc. This allows a comparison with the
shape profiles derived for our BCGs with our deprojection code, which
keeps the direction of principal axes in the 3D deprojected model
fixed.

\begin{table}
\begin{tabular}{ccc}
$\Delta \log \left(\text{M}_\text{tot} / \text{M}_\odot\right)$ & TNG & Magneticum \\
\hline \hline
13-13.35 & 116 & 31\\
$\geq$ 13.35 & 86 & 20\\
\hline
\end{tabular}
\vspace{.5cm}
\caption{
The number of galaxies for every total mass bin from the TNG100 (second column) and Magneticum (third column) simulations.}
\label{Tab.TNG}
\end{table}

\begin{figure*}
\centering
\includegraphics[scale=.3]{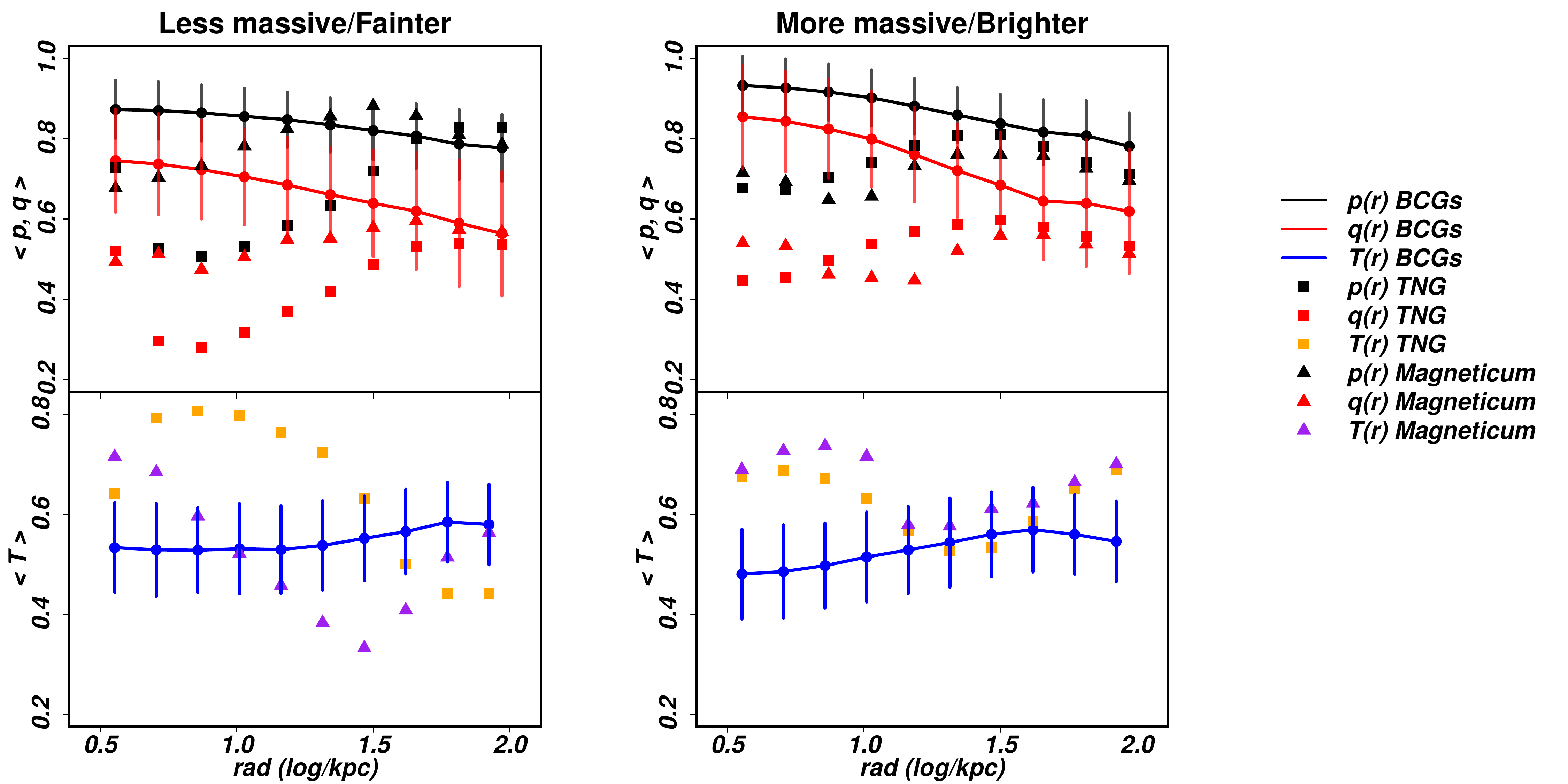}
\caption{Comparison between the $p(r), q(r)$ average profiles (top
  panels) and $T(r)$ (bottom panels) of our BCGs and of the stellar
  component of TNG100 and Magneticum simulations objects. BCGs are
  rendered using lines, while squares and triangles are used for TNG-
  and Magneticum-simulated galaxies, respectively. For the BCGs we
  also compute the RMS in each radial bin, showing it as error bar,
  with a typical value of $\sim 0.08-0.1$.}
\label{Fig.TNG_stars}
\end{figure*}

\begin{figure*}
\centering
\includegraphics[scale=.3]{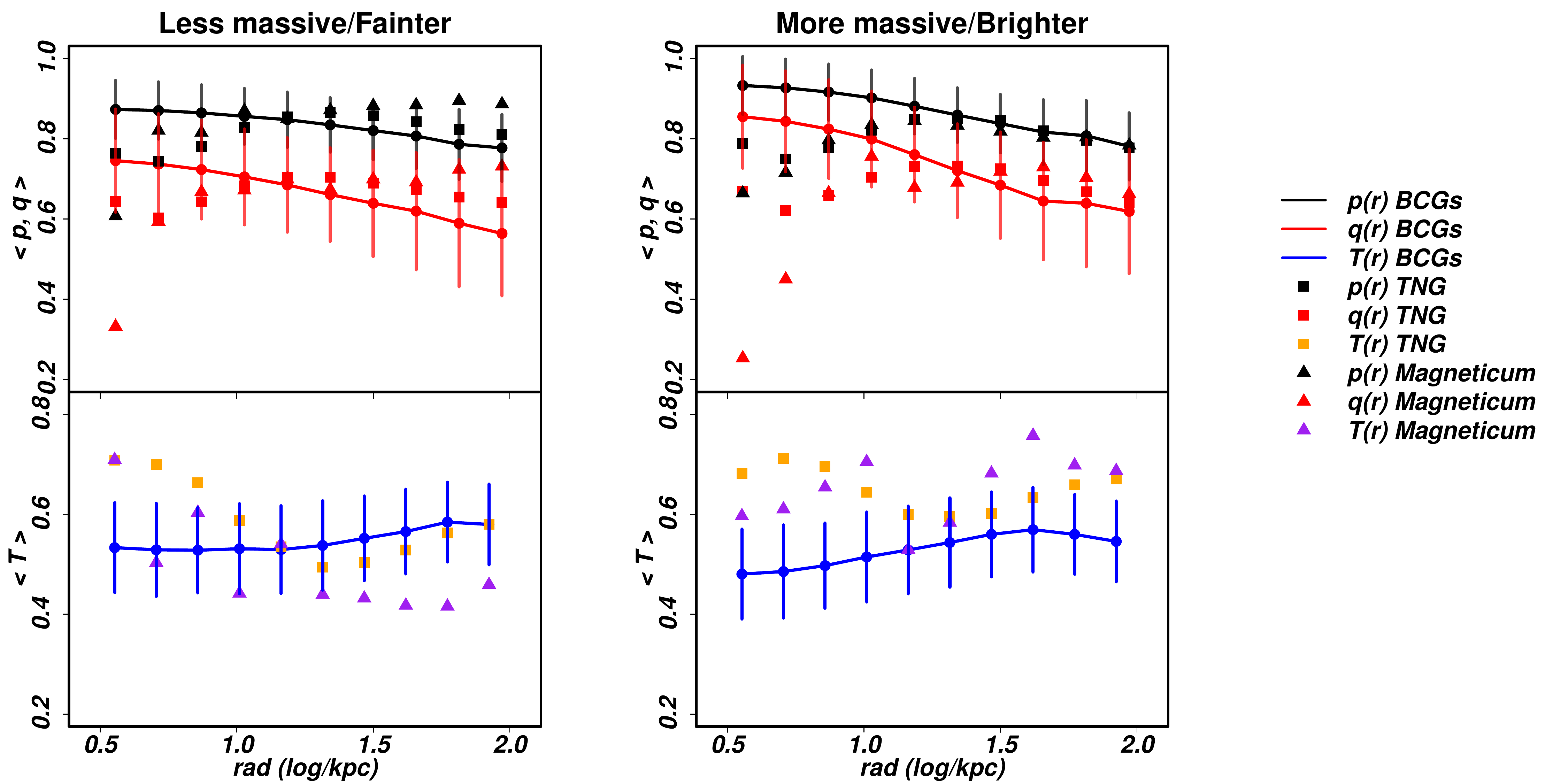}
\caption{Similar to Fig.~\ref{Fig.TNG_stars}, but showing the $p(r),
  q(r)$ and $T(r)$ average profiles of the dark halo component of
  TNG100 and Magneticum simulations objets.}
\label{Fig.TNG_DM}
\end{figure*}

\noindent For each radial bin we compare the average profiles with
those derived for our BCGs with our deprojection code, doing the same
for the triaxiality parameter $T(r)$. We split the BCG sample into a bright one with $\text{M}_{\text{tot}} < -23.7$ and a faint one with $\text{M}_{\text{tot}} > -23.7$, each of which with 21 galaxies. Assuming a M/L ratio of 6, this corresponds to a stellar mass of $2.2 \times 10^{12} M_\odot$. Similarly, we split the simulated galaxies in two samples considering a total mass cut of $2.2 \times 10^{13} M_\odot$. \\
In Figs.~\ref{Fig.TNG_stars} and~\ref{Fig.TNG_DM} we show the comparison between the average
profiles for stars and DM respectively, plotting the BCGs using lines,
the simulated TNG100 galaxies using squares and the simulated
Magneticum galaxies using triangles. For the BCGs we also show the RMS
in each radial bin as error bar, with a typical value of $\sim$0.1. This implies a typical error on the mean value of $\sim$0.02. The RMS for the simulated galaxies is of the same order. The left panels show the faint BCG sample together with the less massive simulated galaxies; the right panels show the bright BCG sample together with the more massive simulated galaxies (see Tab.~\ref{Tab.TNG}).\\
Bright BCGs appear slightly rounder than faint BCGs by $\Delta p \sim $ 0.04 and $\Delta q \sim $ 0.08, but with the same triaxiality. This trend is not obvious when looking at simulated galaxies. \\
The comparison of the profiles of BCGs and simulated galaxies shows that there is a strong disagreement in the
inner regions, especially when the simulated stellar component is considered. The disagreement is less pronounced for the simulated DM halos. In
particular, $p$ and $q$ of TNG100 galaxies have values down to
0.2-0.3, which would imply the presence of squashed structures in almost all
galaxies. The Magneticum galaxies are generally rounder, but still flatter than
the observed ones. This shortcoming of simulations in reproducing the
correct distribution of the ellipticity of massive (slowly rotating)
systems is well documented: different sets of simulations predict a
population of flat slow rotators with ellipticities as high as
0.55-0.6 (\citealt{LiHongyu18} using Illustris\footnote{For these simulations, values as high as 0.8 are found \citep{Claudia20}.}, \citealt{Schulze18}
using Magneticum, \citealt{Naab03} using collision-less $N$-body
simulations). In contrast, the observed ellipticity profiles on the
sky (see top panels of Fig.~\ref{Fig.eps_PA_BCGs}) demonstrate that
most BCGs are round near the center, and it is statistically impossible that
all of them are
axisymmetric systems viewed close to face-on or pole-on. Turning to the
outermost regions, we find a much better agreement between the
profiles of BCGs and of simulated galaxies, for both stars and DM. In
particular, the BCGs and DM $p(r)$ and $q(r)$ profiles from TNG (and
Magneticum at the high-mass end) follow a similar decreasing trend
with a slight offset.

The average profile $T(r)$ of our BCGs is almost flat with $r$ at a
value of $\approx 0.55$ with an RMS scatter of about 0.08, showing that
these objects are overall triaxial. The TNG100 simulation generates
objects which tend to be prolate in the center and as triaxial as our
BCGs in the outer parts. The Magneticum simulation almost matches the
observed average profile at intermediate masses when looking at the
dark component, but produces more oblate/prolate profiles in the
lower/larger mass bins.

Assuming that the flattening of the stellar component in the simulations compared to the observations is due to implementation in the hydrodynamics scheme and that the dark matter is unaffected by this, the similarity between the observed and simulated DM properties and the fact that the observations show similar triaxiality in the outskirts make it plausible that the light distribution of the
outer regions of BCGs is tracing the underlying DM halos and may allow
to even probe the nature of dark matter. In particular, recent
$N$-body simulations (\citealt{Robertson19, Fischer22})
that study mergers of galaxy clusters show that the shape of dark
matter subhaloes depends on their physical properties:
self-interacting dark matter produces rounder halos than classical
$\lambda$CDM. \\

\section{Conclusions}

We have deprojected the photometry of a representative sample of 56
BCGs covering a large radial range with good resolution, from the
innermost to the outermost regions probing into the intracluster
light. The deprojection algorithm is able to generate SB profiles
which are representative of the observed photometry. Moreover, the
results show that the BCGs are consistent with random orientations in
space. For the first time, we have measured radial profiles $p(r),
q(r)$ and $T(r)$. The recovered shapes point to strongly triaxial
galaxies, rounder at the centre and flatter at large radii. A
comparison with the results of the TNG100 and Magneticum simulations shows that BCGs
at large radii are a tracer of the DM halo they are embedded in,
possibly probing the nature of dark matter. Extending this analysis to
galaxies at higher redshifts can probe the
formation history of such objects, although getting SB profiles with
high enough signal-to-noise ratio at large values of $z$ certainly
represents a challenge for present-day facilities. \\ The extremely
strong triaxiality of these objects stresses the need for triaxial
dynamical modeling of the stellar kinematics (e.g. \citealt{Bianca20})
in order to recover unbiased BH mass and M/L estimates, reconstruct
the anisotropy profiles of these galaxies and evaluate the effects of
the different number of DOF. We will address these issues in two
forthcoming papers (de Nicola et. al, in prep.; Neureiter et. al, in
prep.).

\vfill

\section*{Acknowledgements}
We thank the anonymous referee for carefully reading the manuscript and providing us with useful comments
which helped us improving the paper. 
We thank Moritz Fischer for showing us the results of
his PhD thesis previous to submission and for interesting
discussions. \\
The Magneticum Pathfinder simulations were partially performed at the
Leibniz-Rechenzentrum with CPU time assigned to the Project ``pr86re'',
supported by the DFG Cluster of Excellence ``Origin and Structure of the
Universe''. We are especially grateful for the support by M. Petkova
through the Computational Center for Particle and Astrophysics (C2PAP). \\
The LBT is an international collaboration among institutions in the United States, Italy and Germany. LBT Corporation partners are:  LBT Beteiligungsgesellschaft, Germany, representing the Max-Planck Society, the Astrophysical Institute Potsdam, and Heidelberg University; The University of Arizona on behalf of the Arizona university system; Istituto Nazionale di Astrofisica, Italy; The Ohio State University, and The Research Corporation, on behalf of The University of Notre Dame, University of Minnesota and University of Virginia.

\bibliographystyle{aasjournal}
\bibliography{bibl}



\appendix

\begin{figure*}
\subfloat{\includegraphics[width=.5\linewidth]{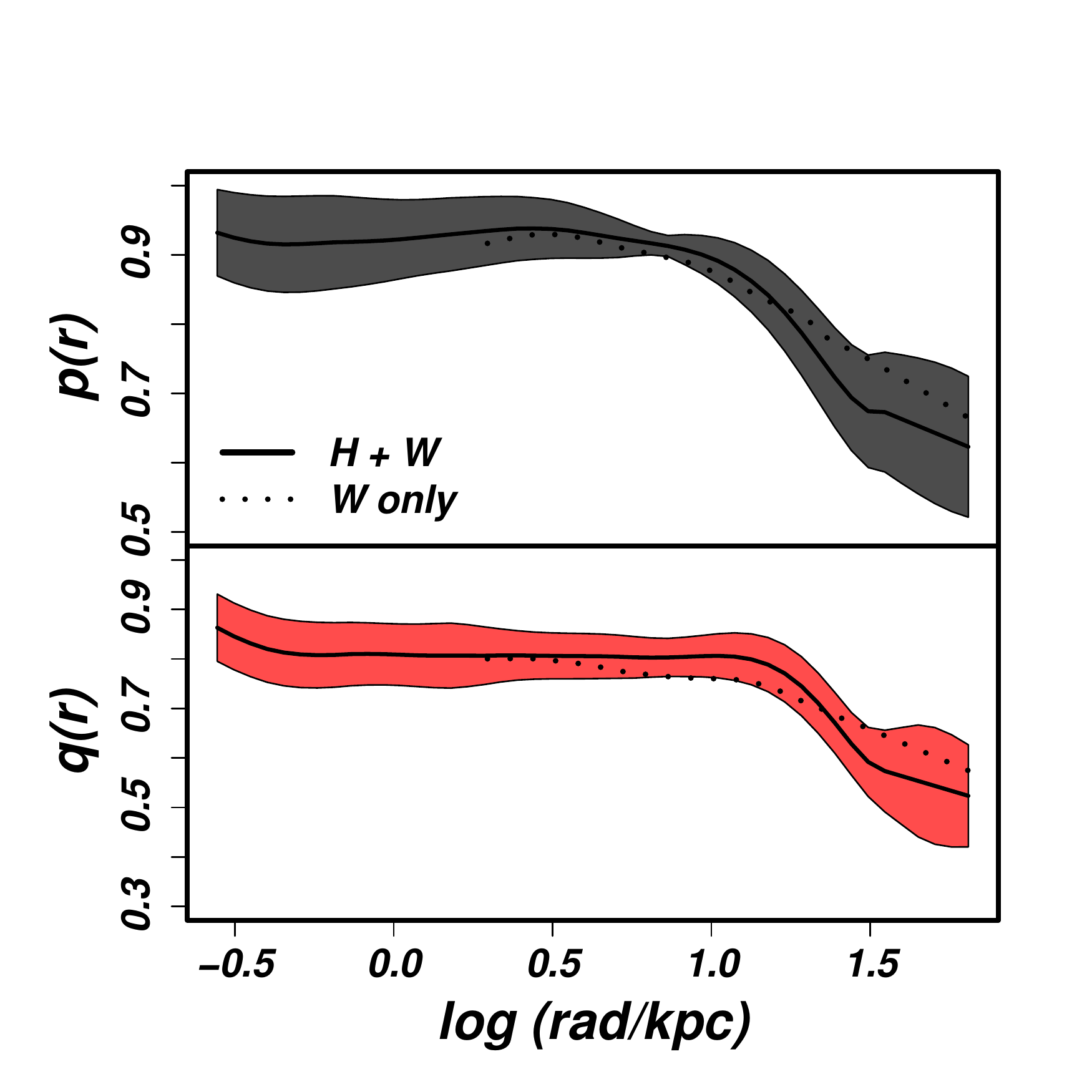}}
\subfloat{\includegraphics[width=.5\linewidth]{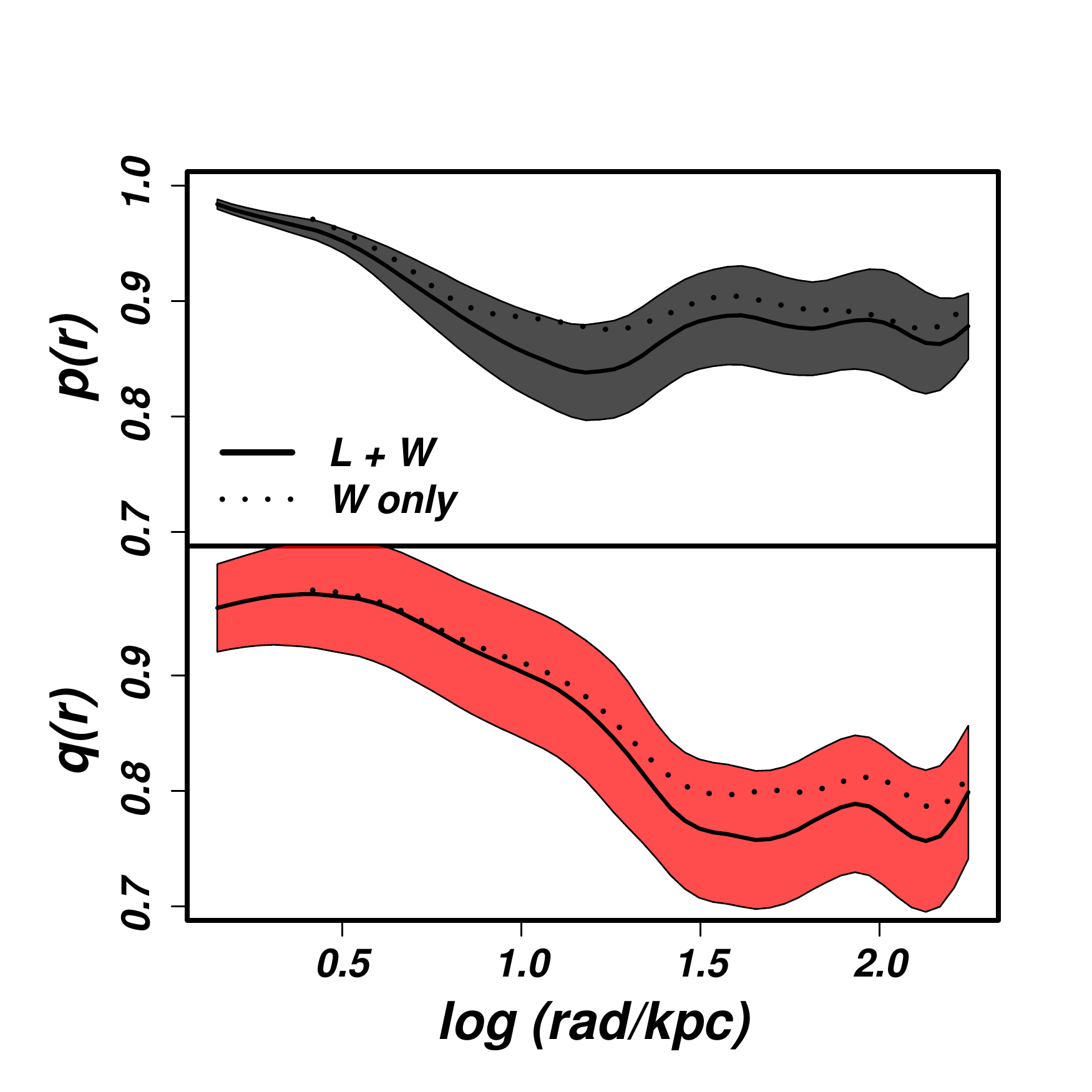}}

\centering
\subfloat{\includegraphics[width=.5\linewidth]{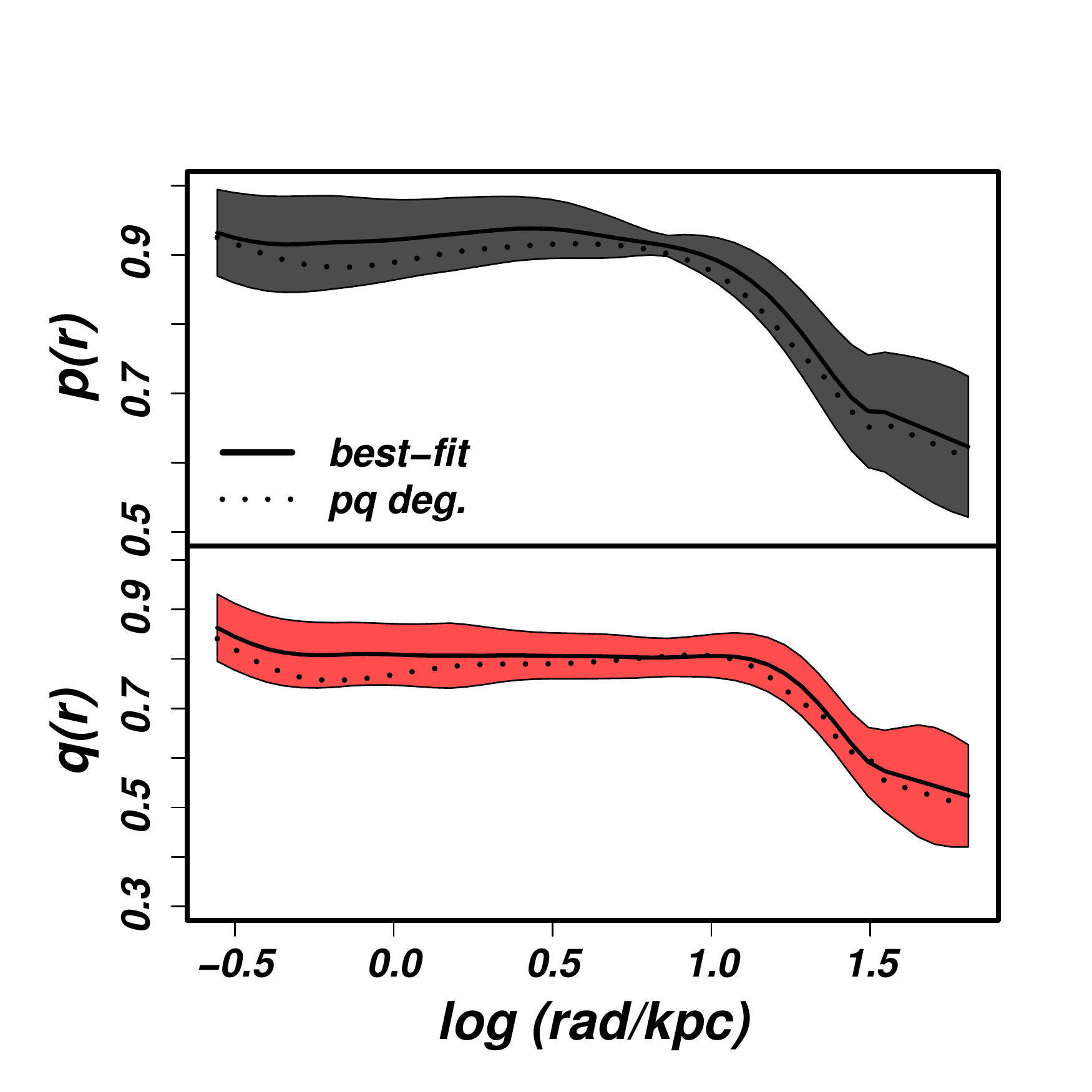}}

\caption{\textit{Top panels:} Comparison of the deprojections obtained at the best-fit viewing angles with or without HST (left, galaxy NGC7647) and LBT photometry (right, galaxy UGC10726). The solid lines are the best-fit profiles with both high-resolution and Wendelstein photometry, while the dotted lines show the solutions with Wendelstein photometry only. The coloured regions are given by the RMS values. \textit{Bottom panel:} For the galaxy NGC7647, we perform a deprojection stopping it when the RMS reaches $1.5 \times \text{RMS}_{\text{min}}$ to judge the effects of possible degeneracies between the model and the viewing angles. The resulting profiles are shown as dotted lines, solid lines and coloured regions are as above.}
    \label{Fig.ref_tests}
\end{figure*}

\section{Resolution and degeneracy effects} \label{App.ref_tests}
We analyze here the systematic effects stemming from the lack of high-resolution HST or LBT data with Wendelstein observations as well as from 
the residual degeneracy between the p \& q profiles and the viewing angles. \\
The first point can easily be investigated by deprojecting galaxies with and without high-resolution data, verifying how much the deprojection differs between the two cases. We choose two galaxies, NGC7647 (for the HST case) and UGC10726 (for the LBT case), and re-perform the deprojection using Wendelstein data only for the best-fit viewing angles. These two galaxies represent stringent tests given the very low scatter for both $p$ and $q$ (see Tab.~\ref{Tab.depro_results}) among different solutions at different viewing angles and the code yielding a low best-fit RMS value. Moreover, the central regions of these galaxies are relaxed, meaning that we can exploit HST and LBT data up to the innermost radii. \\
In the two top panels of Fig.~\ref{Fig.ref_tests} we show the $p(r), q(r)$ profiles from the HST(LBT)+Wendelstein case as solid lines, while the Wendelstein-only profiles are shown as dotted lines for $p$ and $q$. The Wendelstein-only deprojection cannot probe the inner region of the galaxy, but remains within the region delimited by the RMS (shown as coloured area in the figures) at larger radii.  \\
As a second test, we take the galaxy NGC7647 and deproject it at the best-fit viewing angles (with HST photometry) stopping the deprojection as soon as the RMS reaches $1.5 \times \text{RMS}_{\text{min}}$. In the bottom panel of Fig.~\ref{Fig.ref_tests} we compare this solution to the best-fit one. Also in this case the deviations from the best-fit solution are smaller than the scatter due to the different viewing angles for which an acceptable deprojection is found. \\
Thus, we conclude the lack of the high-resolution photometry does not change the conclusions reported in this paper. Moreover,  considering only the best-fit solution (in terms of the RMS) for a given set of viewing angles probes the range of acceptable p \& q profiles.

\section{Ellipticity and PA profiles} \label{App.profiles}

\begin{figure}
\centering
\includegraphics[scale=.5]{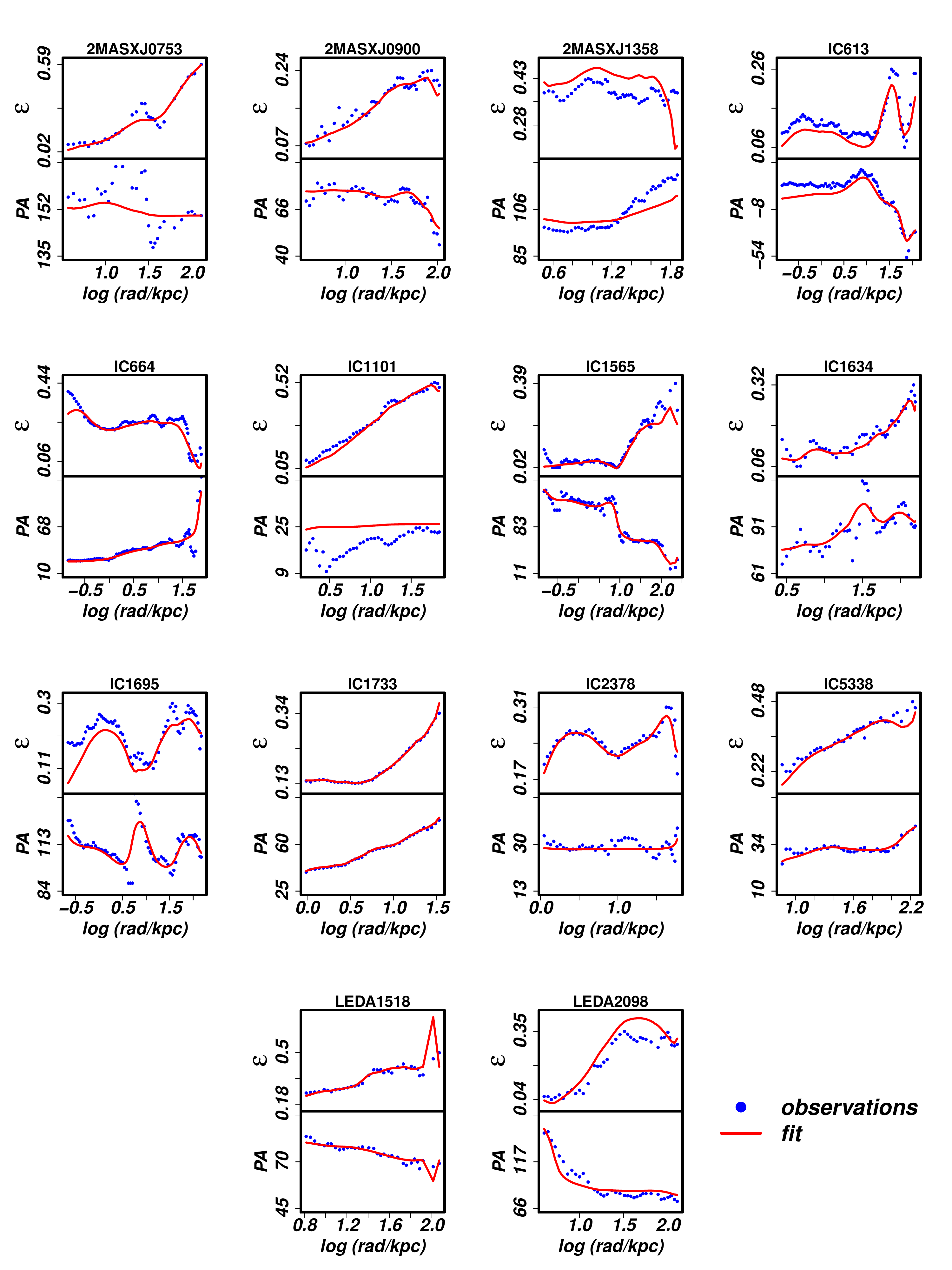}
\caption{\ellip (top panels) and PA (bottom panels) profiles of every
  BCG considered in this work. Blue points are the original
  photometry, whereas the red lines show our fits. The radii are given
  in log$_{10}$ kpc.}
\label{Fig.eps_PA_BCGs}
\end{figure}

In Fig.~\ref{Fig.eps_PA_BCGs} we show the $\varepsilon$ and the PA
profiles for the BCGs of our sample. Omitted points (see also notes in
App.~\ref{App.notes}) are not shown.

\begin{figure}
\centering
\includegraphics[scale=.5]{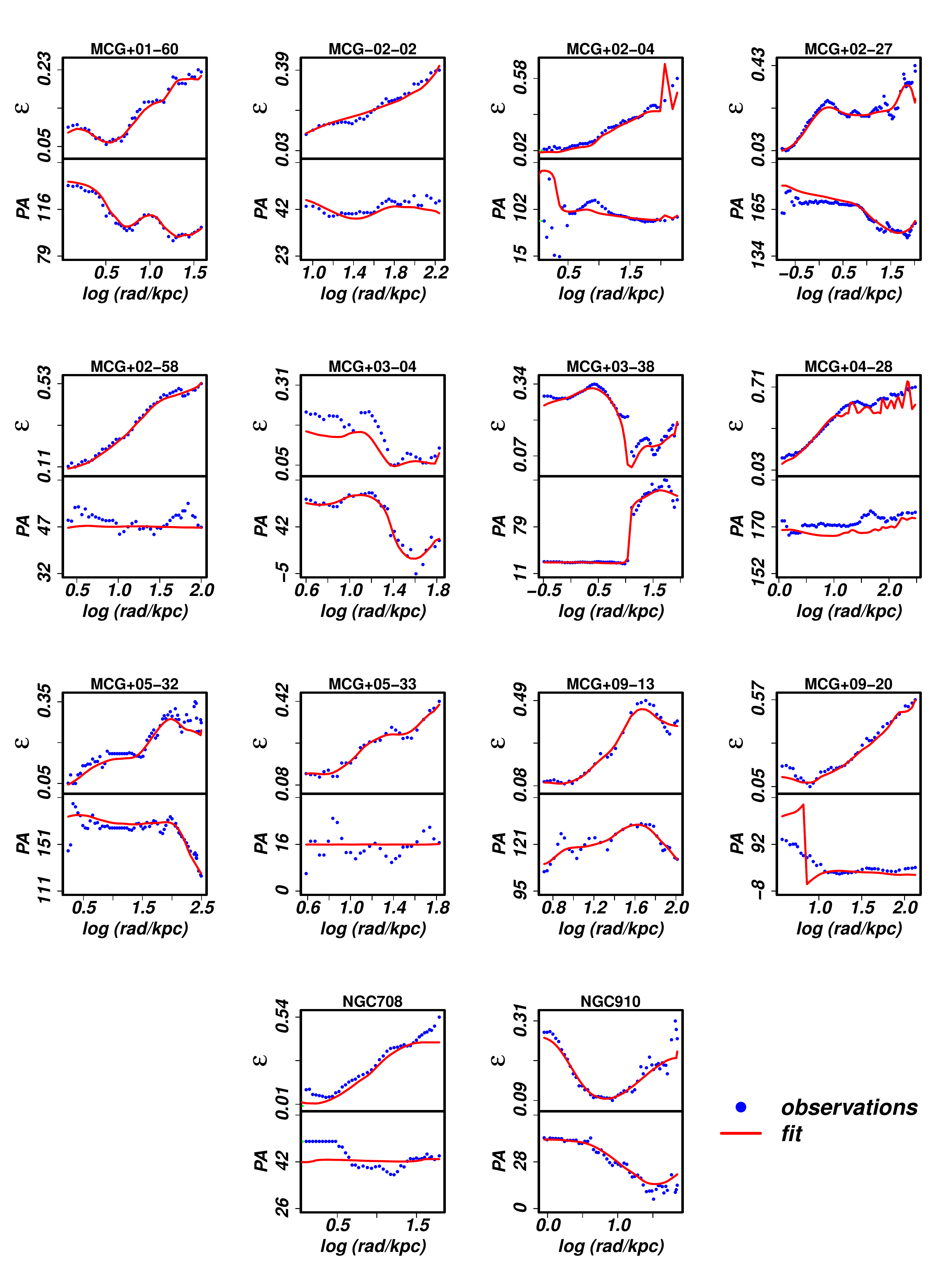}
\caption*{Figure 8 \textit{(continued)}}
\label{Fig.eps_PA_BCGs_p2}
\end{figure}

\begin{figure}
\centering
\includegraphics[scale=.5]{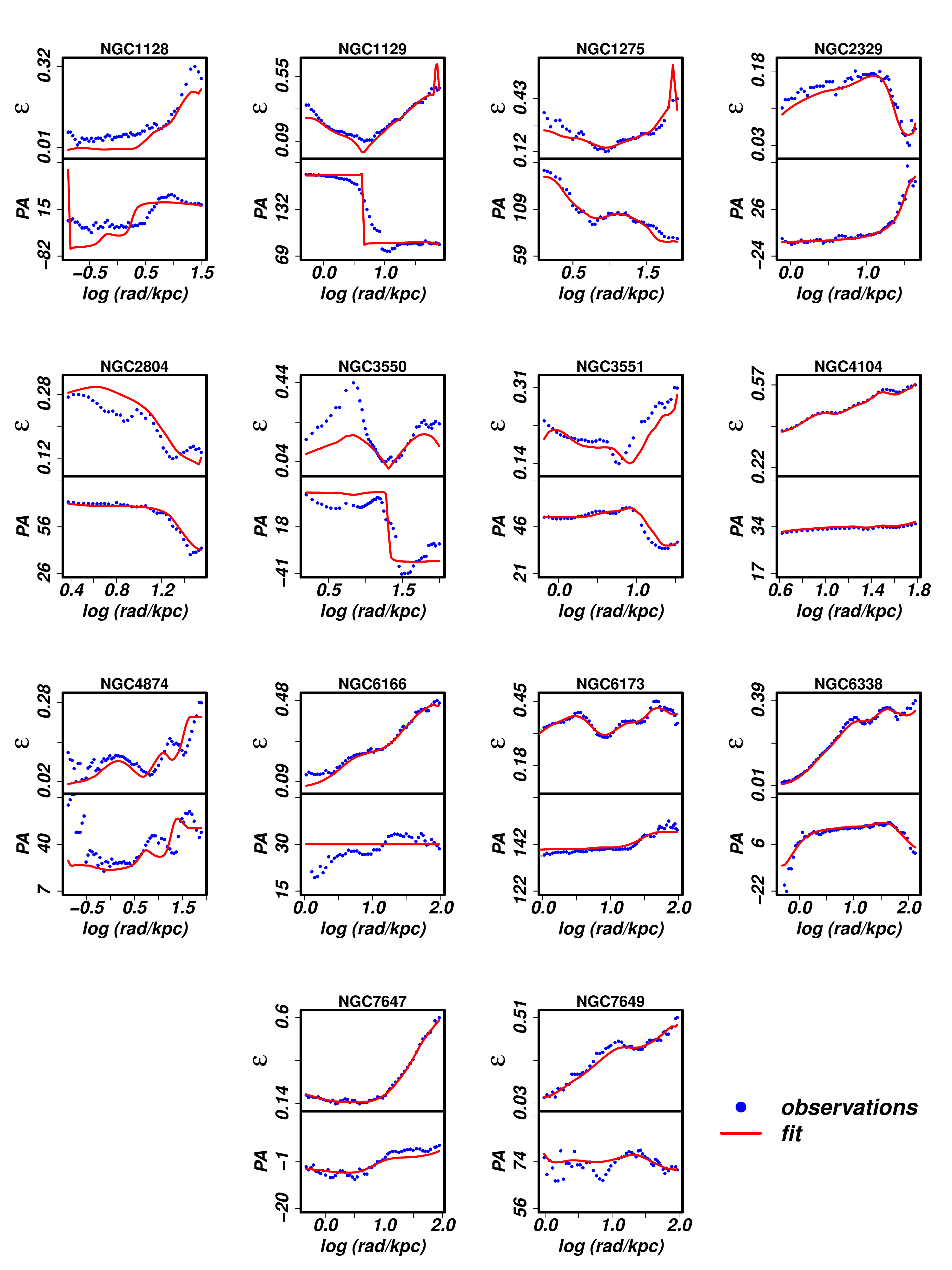}
\caption*{Figure 8 \textit{(continued)}}
\label{Fig.eps_PA_BCGs_p3}
\end{figure}

\begin{figure}
\centering
\includegraphics[scale=.5]{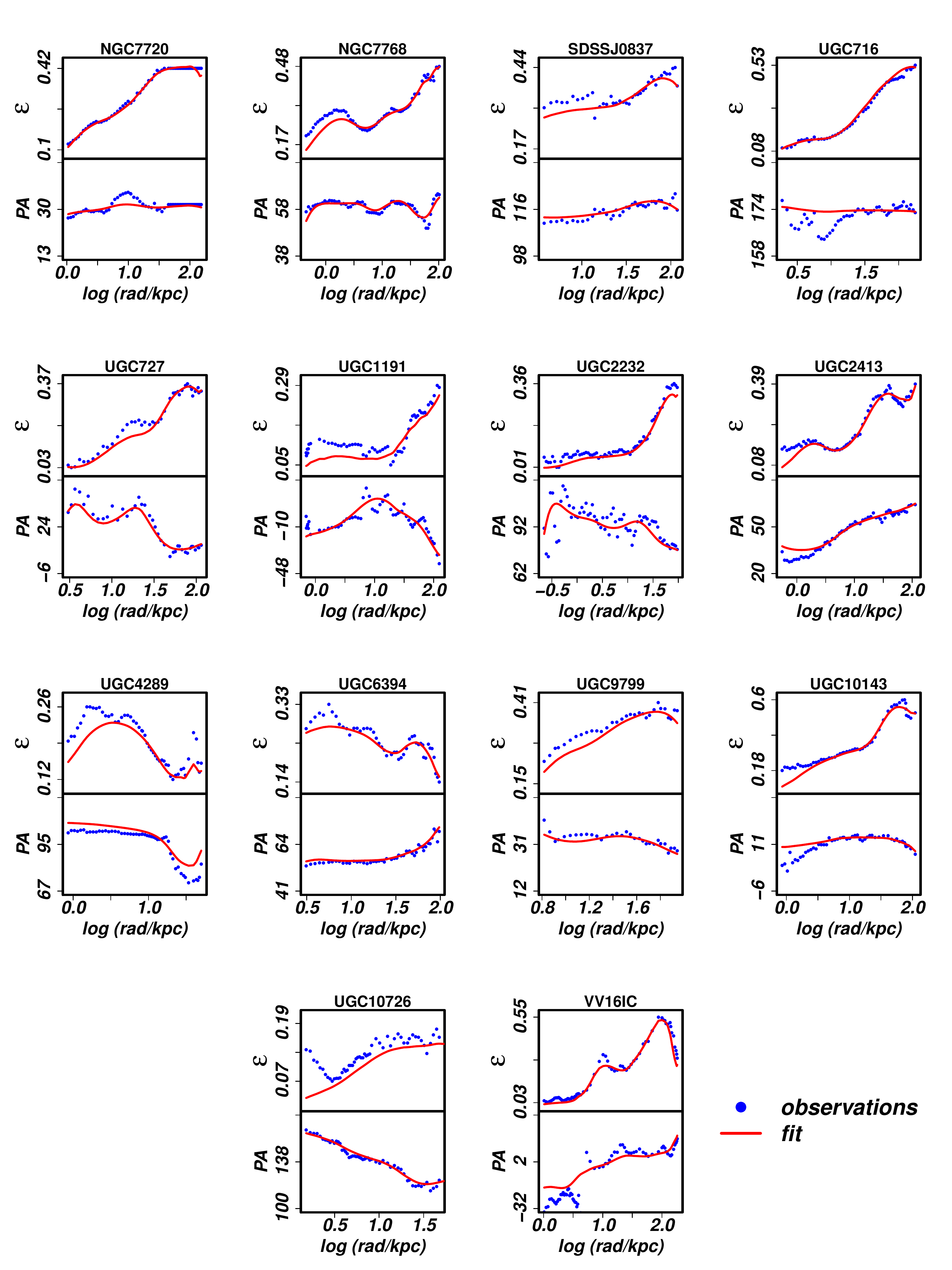}
\caption*{Figure 8 \textit{(continued)}}
\label{Fig.eps_PA_BCGs_p4}
\end{figure}


\section{Intrinsic shape profiles} \label{App.intr_profiles}

\begin{figure}
\centering
\includegraphics[scale=.5]{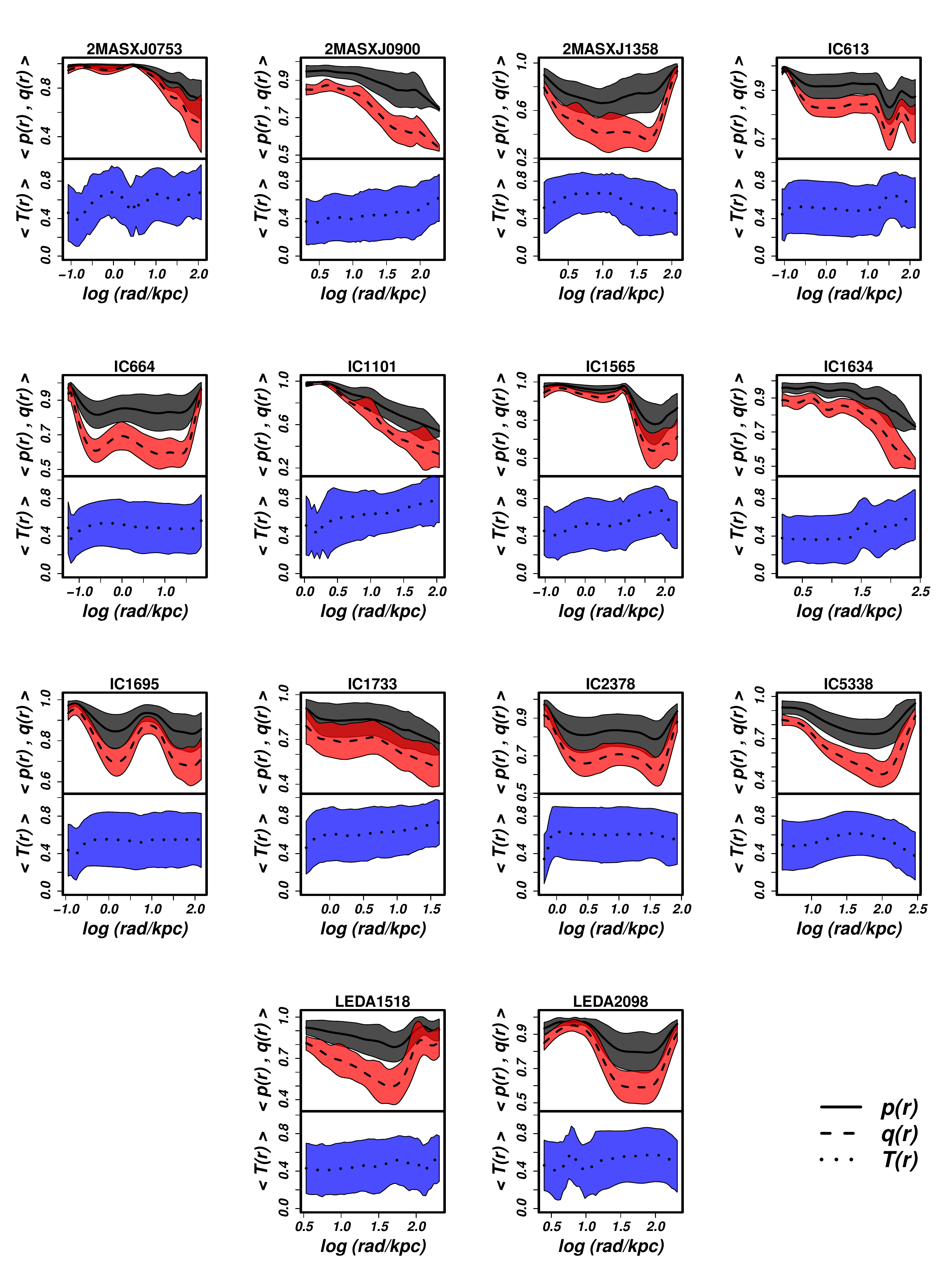}
\caption{Intrinsic axis ratio $p, q$ (top panels) and triaxiality $T$ (bottom panels) profiles of every BCG considered in this work. The solid, dashed and dotted lines are the average among all good profiles (see Sec.~\ref{Ssec.va}), while the coloured regions mark the RMS values. The radii are given in log$_{10}$ kpc.}
\label{Fig.pqT_BCGs}
\end{figure}

In Fig.~\ref{Fig.pqT_BCGs} we show intrinsic axis ratio profiles, along with the corresponding triaxiality profiles, for every BCG of the sample. The profiles are computed by averaging over all acceptable deprojections.

\begin{figure}
\centering
\includegraphics[scale=.5]{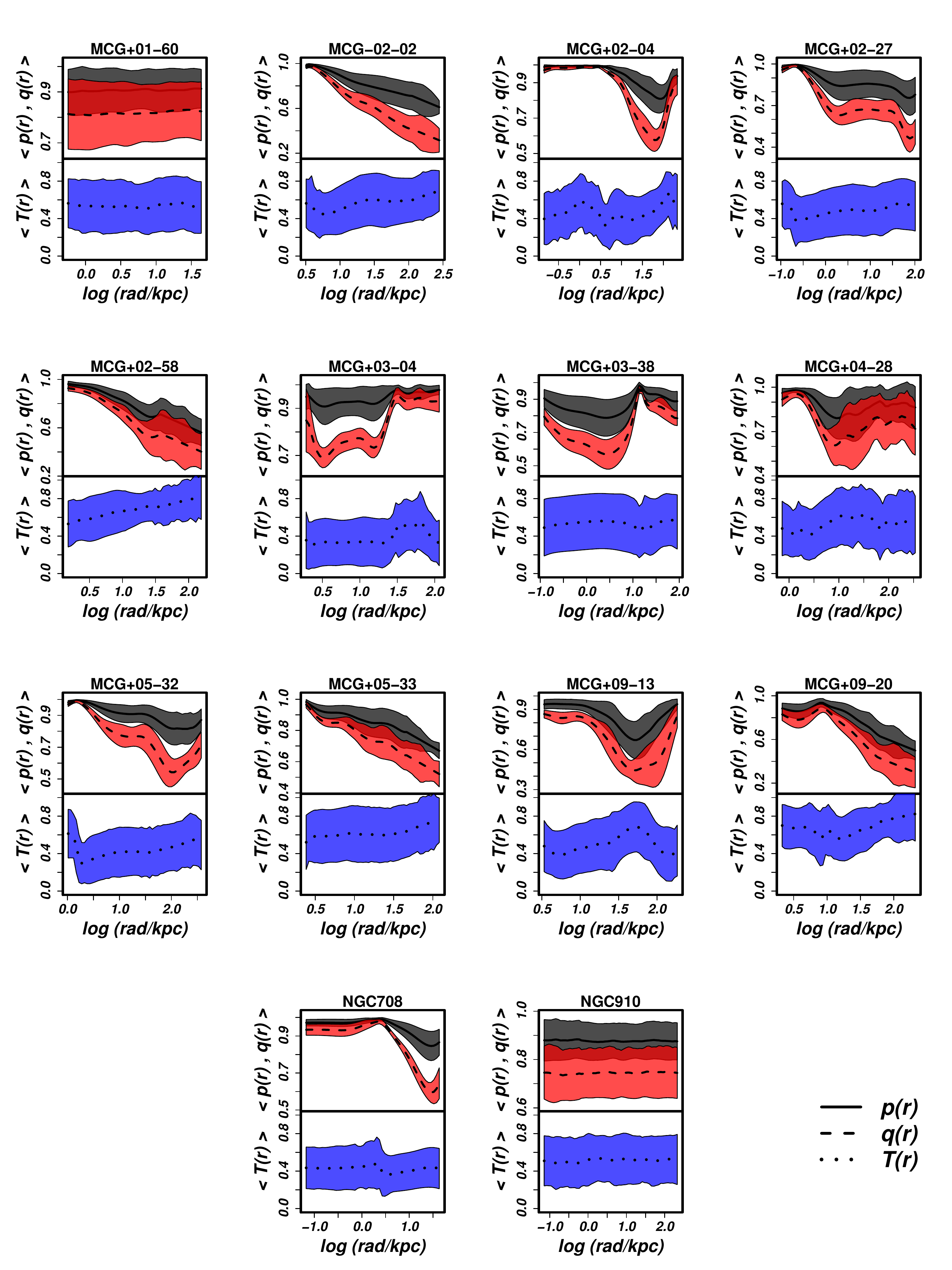}
\caption*{Figure 9 \textit{(continued)}}
\label{Fig.pqT_BCGs_p2}
\end{figure}

\begin{figure}
\centering
\includegraphics[scale=.5]{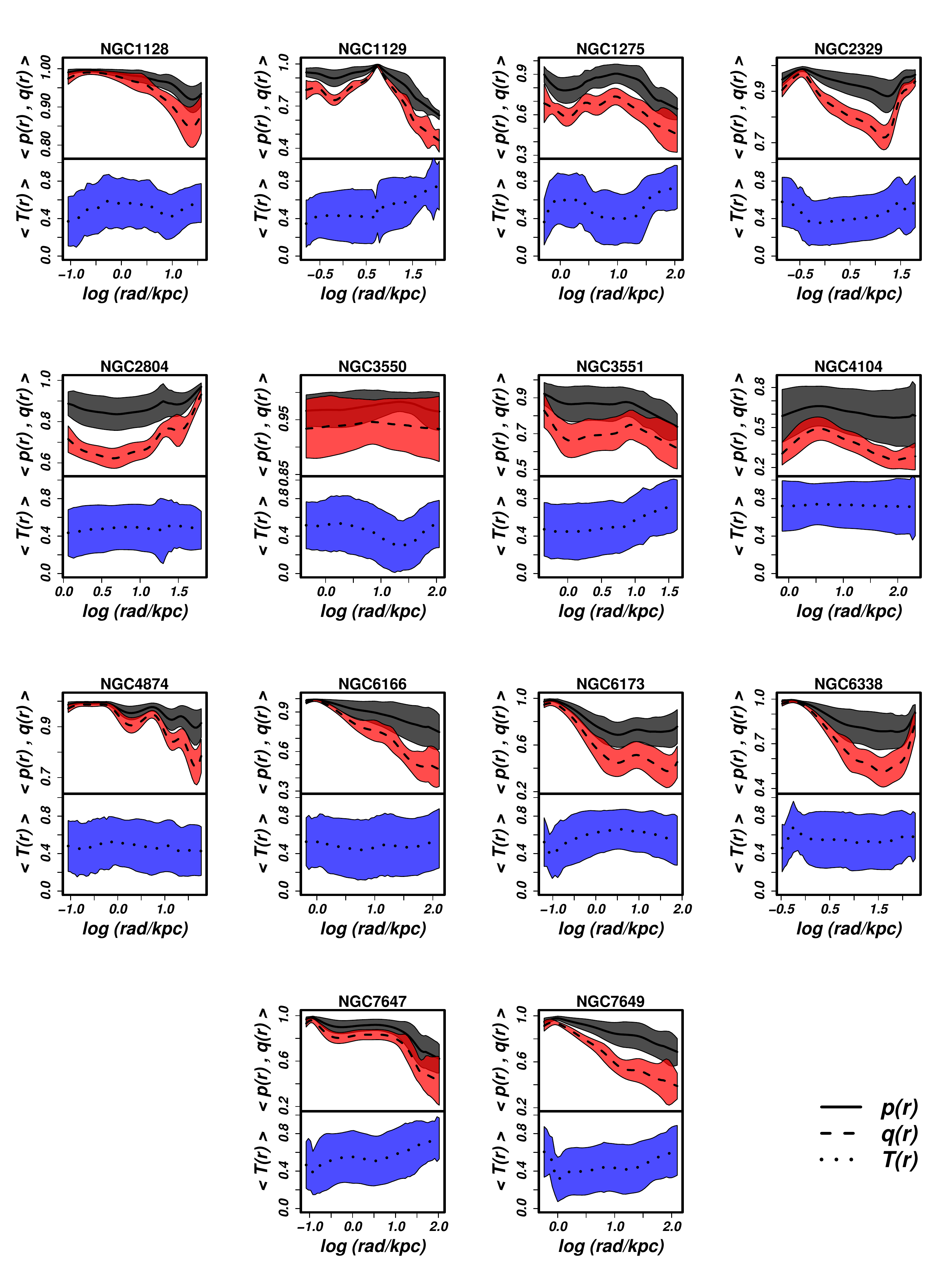}
\caption*{Figure 9 \textit{(continued)}}
\label{Fig.pqT_BCGs_p3}
\end{figure}

\begin{figure}
\centering
\includegraphics[scale=.5]{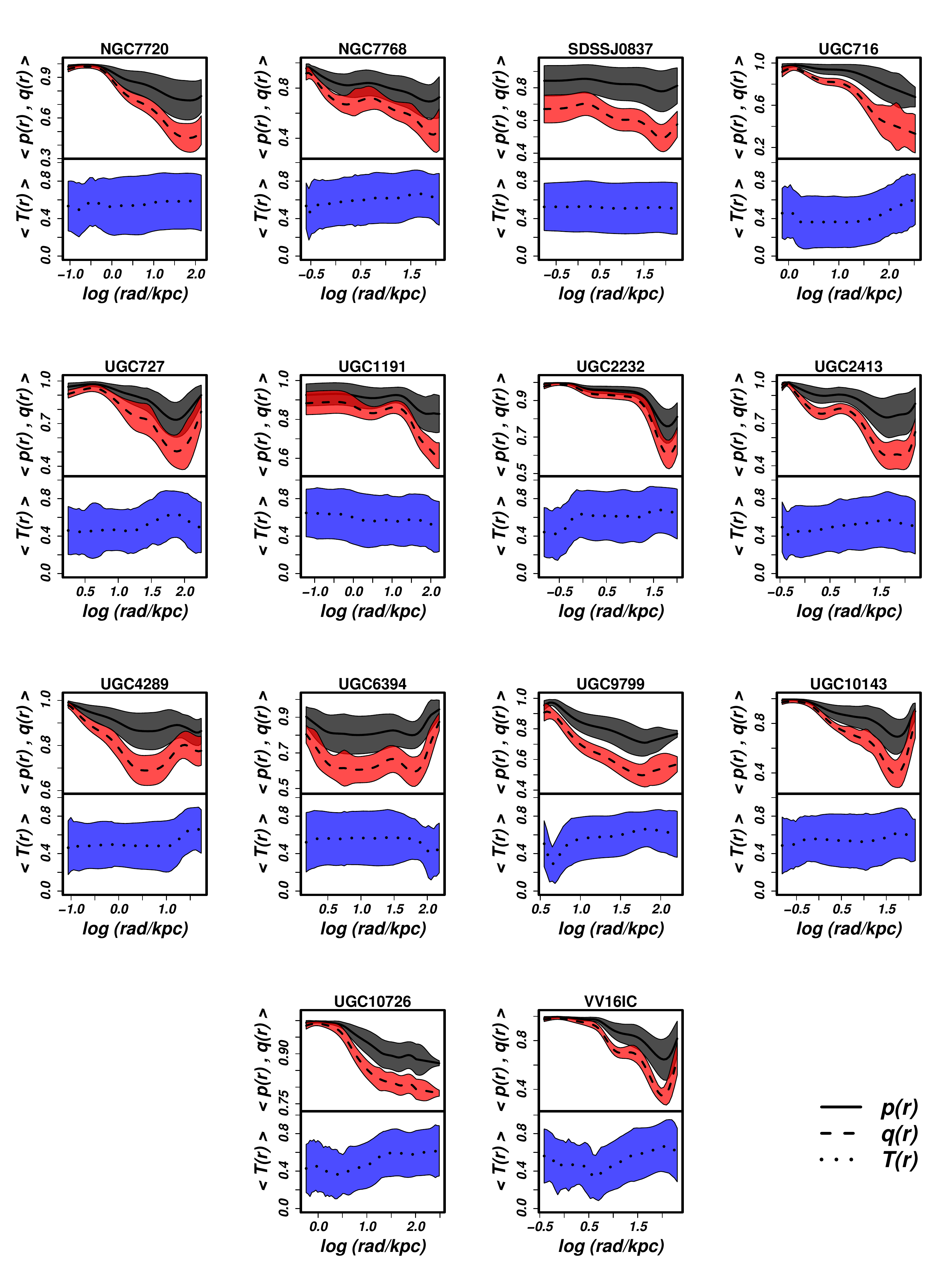}
\caption*{Figure 9 \textit{(continued)}}
\label{Fig.pqT_BCGs_p4}
\end{figure}

\section{Notes on individual galaxies} \label{App.notes}

\begin{itemize}

    \item \textbf{2MASXJ0753:} This is one of the galaxies observed at LBT. The PSF effects are clearly visible way beyond the 0.4" seeing value estimated during the observations, and therefore all the affected points are not taken into account. The galaxy shows bumpy/noisy \ellip and PA profiles which cannot be described accurately, despite the fact that the viewing angles are not close to the principal axes.

    \item \textbf{2MASXJ0900:} This well-fitted galaxy has a sudden 25$^\circ$ twist in the outermost regions, which is well reproduced. This may be due to the ICL given that in the innermost the regions there is no significant twist.
 
    \item \textbf{2MASXJ1358:} This is an example of a galaxy whose \ellip does not change much as a function of the radius. Our best-fit slightly overestimates it, while the twist is underestimated. We stop the deprojection at 75 kpc since the isophotal parameters cannot be adequately fitted anymore beyond this radius.

    \item \textbf{IC613:} The galaxy is very round (\ellip $\lesssim$ 0.1 until beyond 10 kpc). The outermost radii hint at something not in equilibrium, which the code does fit well. The PA at the centre is not well reproduced, but since \ellip is small in the central region, this is not a serious issue.
    
    \item \textbf{IC664:} This is a rare example of a galaxy which is flat in the center and round in the outskirts. This low \ellip generates an unrealistic twist at large radii, but the galaxy still shows a nice constant PA at lower radii which is well reproduced by the code.

    \item \textbf{IC1101:} The best-fit angles \angles = (50,80,160)\grad\, almost lie on the $(y, z)$ plane, which does not allow for a precise recovery of the twist. Possibly these angles are not the correct ones, because one of the random re-projections does produce a better fit to the observed twist. The somewhat bumpy \ellip and PA profiles point to the presence of not fully relaxed structures. 
    
    \item \textbf{IC1565:} The galaxy is well fitted with the exception of the outermost \ellip points which likely belong to the intracluster light. The huge twist shows some oscillations, hinting at a not completely relaxed galaxy. 
    
    \item \textbf{IC1634:} The galaxy has very noisy profiles, which hint at a not yet relaxed galaxy. However, the code reproduces the trends very well. 
    
    \item \textbf{IC1695:} Although both the \ellip and the PA profiles are complex, the code reproduces these profiles well, with the exception of the major bump in the PA, which is underestimated by a factor of 2. 
    
    \item \textbf{IC1733:} We stop the deprojection at 35 kpc because of possible unrelaxed structures at larger radii. It is the best-fitted galaxy of the sample, with an RMS of 0.0073.
    
    \item \textbf{IC2378:} The best-fit angles \angles = (70,10,150)\grad\, are close to the $x$-axis. The somewhat noisy but almost constant PA profile allows a good fit with low RMS and indicates that the galaxy might be oriented along one of the principal axes. 
    
    \item \textbf{IC5338:} We start the deprojection at 4.9 kpc because of possible AGN contamination. The last points do give a suspicious twist increase probably linked to the intracluster light; however, this is nicely reproduced.
    
    \item \textbf{LEDA1518:} The galaxy has a 20$^\circ$ twist and a smooth ellipticity profile, although we note that \ellip $>$ 0.2 in the central regions already. The deprojection reproduces the profiles very well except for the outermost points.
    
    \item \textbf{LEDA2098:} The galaxy has a huge 80$^\circ$ twist, but this is mostly given by the very low ellipticity in the central regions. The code reproduces these profiles well, slightly overestimating \ellip at the outer radii.

    \item \textbf{MCG+01-60:} We start the deprojection at 1.16 kpc, since for this galaxy only Wendelstein data are available. This is another galaxy with $\text{RMS}_\text{best} \leq 0.01$.
    
    \item \textbf{MCG-02-02:} This well relaxed galaxy has a typical ellipticity profile rising steadily and a small 10$^\circ$ twist. No solutions compatible with v.a. along the principal axes are found, but we get acceptable deprojections at $\theta = 10^\circ$.
    
    \item \textbf{MCG+02-04:} The twist is well reproduced with the exception of a bump around 10 kpc. In the central regions we suspect the galaxy to be not fully relaxed, because of an unrealistic $\sim$150\grad twist at small radii. Therefore, we start the deprojection at 1.3 kpc.

    \item \textbf{MCG+02-27:} The 30\grad\,twist present in the central regions is not reproduced very well; however since \ellip is very low this does not significantly affect the goodness of the fit. The bumpy \ellip profile suggests that relaxation is not complete.

    \item \textbf{MCG+02-58, MCG+05-33:} The best-fit angles \angles = (80,0,165)\grad\,and \angles = (60,10,130)\grad\,almost lie along the $x$-axis and on the $(x, z)$ plane, respectively. This does not allow for a good twist recovery; however, the true PA profiles oscillate around 15\grad\, (which would indeed give $\psi = 165^\circ$ if the galaxy were along $x$) for MCG+05-33 and around 50\grad\, for MCG+02-58 (for which $\psi = 130^\circ$ would be the right value). We measure twists oscillating around 20\grad and 9\grad, respectively. This indicates that the two galaxies are oriented along one of the principal axes, but possibly not fully relaxed yet.

    \item \textbf{MCG+03-04:} This interesting galaxy is flat in the central regions, where our code slightly underestimates the ellipticity, and gets rounder in the outskirts. The PA profile is tricky, since the twist is small in the central regions before getting significantly bigger at large radii, where \ellip is small. The fact that the twist is small in the central regions enables the code to obtain good fits close to the principal axes, as for IC2378.

    \item \textbf{MCG+03-38:} Another flat galaxy in the central regions. Both \ellip and PA jump wildly at $\sim$10-15 kpc, as if a decoupled structure were there in the centre. Nevertheless, the code returns a good fit to the observed photometry.

    \item \textbf{MCG+04-28:} The high RMS (0.122) indicates that the fit is not satisfactory beyond $\sim$10 kpc, where the galaxy becomes extremely flat. The somewhat bumpy twist is overall small, therefore the systematic offset between model and data is not worrying.

    \item \textbf{MCG+05-32:} The galaxy is well fitted. The photometry shows an unrealistic twist in the first $\sim$10 isophotes, which is probably the result of the low ellipticity.

    \item \textbf{MCG+09-13, MCG+09-20:} These are both galaxies showing typical isophotes of massive ellipticals, although with some bumps. The twist in the central regions for MCG+09-20 is due to \ellip almost going to 0.

    \item \textbf{NGC708:} Although the best-fit viewing angles are not exactly on the principal axes, there are several good solutions compatible with such inclinations, as shown in Fig.~\ref{Fig.angles_octant}. The photometry has not been deprojected within the first 1.2 kpc because of a dust lane. The scale of the plot in Fig.~\ref{Fig.eps_PA_BCGs} might give the wrong impression of a poorly recovered twist, which is not the case.

    \item \textbf{NGC910:} Like NGC1129, \ellip goes down and then up again. The 30\grad\, twist is well fitted.

    \item \textbf{NGC1128:} We do not include the galaxy in the twist histogram, as \ellip is almost always below 0.1, except for the outermost radii.
    
    \item \textbf{NGC1129:} The best-fit angles \angles = (60,10,0)\grad\, almost lie on the $(x, z)$ plane. \ellip goes down to 0 and then increases again. Given that the twist is roughly 90\grad, this could be a galaxy compatible with viewing angles along the principal axes (as suggested by the best-fit viewing angles) \textit{and} with intersecting $p, q$ profiles.
    
    \item \textbf{NGC1275:} This peculiar galaxy shows a high ellipticity both in the innermost and in the outermost regions, with a dip in between. The 80$^\circ$ twist is very well recovered.
    
    \item \textbf{NGC2329:} We omit the innermost points because of an unrealistic bump in \ellip.

    \item \textbf{NGC2804:} \ellip decreases towards the outermost regions. In the first 10 kpc, the twist is completely absent. We stop the deprojection at 65 kpc because the isophotal parameters cannot be measured anymore beyond this radius.
    
    \item \textbf{NGC3550:} This galaxy was observed at LBT under poor seeing conditions. Moreover, both the \ellip and the PA profiles hint at a not fully relaxed galaxy. 
    
    \item \textbf{NCG3551:} We omit the innermost three isophotes because of resolution problems when deriving the isophotal parameters (only Wendelstein images are available for this galaxy). The deprojection beyond 35 kpc also becomes unfeasible since the galaxy shows signs of non-equilibrium, however the deprojection yields RMS $\leq$ 0.01.
    
    \item \textbf{NGC4104:} The first 1.5" arcseconds must be discarded because of poor seeing. Also the outermost points (from 60 kpc) are omitted due to contamination from a neighbor galaxy. It is one of the flattest galaxies of the sample, with \ellip always between 0.4-0.6.
    
     \item \textbf{NGC4874:} Very round galaxy with noisy profiles. We include it in the histogram, although the only region where \ellip stabilizes above 0.1 is beyond 10 kpc. 
     
     \item \textbf{NGC6166:} The best-fit angles \angles = (40,0,60)\grad\, lie on the $(x, z)$ plane. The low RMS might be explained by the fact that with the exception of the innermost radii (where most of the twist occurs but \ellip is low) the true PA oscillates around the constant PA recovered by the code.
     
     \item \textbf{NGC6173:} We dropped the poorly fitted central region of the galaxy. 
     
     \item \textbf{NGC6338:} The deprojection starts at $\sim$0.6" because of possible AGN activity.  
     
     \item \textbf{NGC7647:} Nothing to signal here.
     
     \item \textbf{NGC7649, NGC7720, NGC7768:} It is not entirely clear how relaxed the galaxies are. The PA profiles are somewhat noisy with very small twists, while the \ellip profiles increase smoothly with radius (with the exception of the central regions of NGC7768) with minor dips.

     \item \textbf{SDSSJ0837:} The same considerations about the observations made for 2MASXJ0837 also apply for this galaxy. However, this galaxy does not show signs of non-equilibrium.
     
     \item \textbf{UGC716:} The best-fit angles \angles = (40,0,60)\grad\, almost lie on the $(y, z)$ plane. This is another galaxy which might indeed be close to the principal axes despite the small $\sim$10\grad\, twist near the round center.
     
     \item \textbf{UGC727:} See comments of MCG+09-13 and MCG+09-20.

     \item \textbf{UGC1191:} We start the deprojection at $\sim$0.6-0.7 kpc because of PSF effects. The galaxy has  a large twist ($\sim$40\grad) which is well reproduced. 
     
     \item \textbf{UGC2232:} Nothing to signal here.
     
     \item \textbf{UGC2413:} We note a slight offset in the central regions between the true photometry and the recovered one, probably because of resolution effects given by the spherical $\rho$-grid.
     
     \item \textbf{UGC4289:} The galaxy shows a somewhat noisy \ellip profile along with a PA profile with an abrupt $\sim$30\grad\, twist starting from $\sim$20 kpc. We stop the deprojection at 50 kpc because of possible contamination from neighbour galaxies.

     \item \textbf{UGC6394:} The ellipticity decreases as a function of radius and shows some bumps, while the PA profile is much smoother and very well recovered. For this galaxy we also obtain prolate deprojections compatible with the observed photometry.

     \item \textbf{UGC9799:} We start the deprojection at 4 kpc to avoid the center affected by probable AGN contamination.
    
     \item \textbf{UGC10143:} The same considerations made for UGC2413 also apply to this galaxy. We start the deprojection at 0.6 kpc because of a sudden 100\grad\,twist in the innermost regions. 
     
     \item \textbf{UGC10726:} The same considerations made for UGC1191 also apply to this galaxy. We omit the outermost isophotes.

     \item \textbf{VV16IC:} See comments of UGC9799.
    
\end{itemize}

\end{document}